\documentclass[journal]{IEEEtran}
\usepackage{amsmath,amsfonts,amssymb,mathtools}
\usepackage{url}
\usepackage{graphicx}
\usepackage{cite,color,multirow}
\usepackage{stix}
\usepackage{tikz}
\usetikzlibrary{matrix}
\usetikzlibrary{positioning}
\usetikzlibrary{decorations.pathreplacing}

\def\ve#1{{\mathchoice{\mbox{\boldmath$\displaystyle #1$}}%
{\mbox{\boldmath$\textstyle #1$}}%
{\mbox{\boldmath$\scriptstyle #1$}}%
{\mbox{\boldmath$\scriptscriptstyle #1$}}}}
\def\j{\mathrm{j}}
\def\e{\mathrm{e}}

\def\x{\mathrm{x}}

\def\munsh{\mu_n^{\mathrm{sh}}}
\def\mufsh{\mu_4^{\mathrm{sh}}}
\def\mununi{\mu_n^{\mathrm{uni}}}
\def\nush{{\nu_{\mathrm{sh}}}}
\def\etash{{\eta_{\mathrm{sh}}}}
\def\nuref{{\nu_{\mathrm{ref}}}}
\def\etaref{{\eta_{\mathrm{ref}}}}
\def\pase{P_{\mathrm{ASE}}}
\def\CS{M} 
\def\Pnlin{P_{\mathrm{NLIN}}}
\def\Popt{P_{\mathrm{opt}}}
\def\dminsq{d^2_{\mathrm{min}}}
\def\Qmaxsh{{Q_{\mathrm{max},\mathrm{sh}}}}
\def\Qmaxref{{Q_{\mathrm{max, ref}}}}
\def\SNRmaxsh{{\mathrm{SNR_{eff,sh}^{opt}}}}\def\SNRmaxref{{\mathrm{SNR_{eff,ref}^{opt}}}}

\def\Poptsh{{P^{\mathrm{opt}}_{\mathrm{sh}}}}
\def\Poptref{{P^{\mathrm{opt}}_{\mathrm{ref}}}}
\def\Qmax{{Q_{\mathrm{max}}}}
\def\SNRnlopt{\mathrm{SNR_{eff}^{opt}}}
\def\SNRnl{\mathrm{SNR_{eff}}}
\def\glin{g_\mathrm{lin}}
\def\gnlin{g_\mathrm{nl}}
\def\Bch{B_{\mathrm{ch}}}
\def\Rs{R_{\mathrm{s}}}
\def\Rshaper{R_{\mathrm{sh}}}
\def\Rloss{R_{\mathrm{loss}}}
\def\wspm{w_{\mathrm{spm}}}
\def\wxpm{w_{\mathrm{xpm}}}
\def\wmem{w_{\mathrm{mem}}}
\def\Rs{R_{\mathrm{s}}}
\def\Nch{N_{\mathrm{ch}}}

\def\edp{e_{\mathrm{x/y}}}
\def\Ncpr{N_{\mathrm{cpr}}}
\def\t{t}
\def\r{r_{\mathrm{a}}}
\def\m{m}
\def\metricam{m_{\mathrm{AM}}}
\def\Npb{N_{\mathrm{pb}}}
\def\CoAM{C_{\mathrm{AM}}}

\newcommand{\appropto}{\mathrel{\vcenter{
  \offinterlineskip\halign{\hfil$##$\cr
    \propto\cr\noalign{\kern2pt}\sim\cr\noalign{\kern-2pt}}}}}

\begin{document}

\title{Probabilistic Shaping for Nonlinearity Tolerance}

\author{Mohammad Taha Askari and Lutz Lampe,~\IEEEmembership{Senior Member,~IEEE}
\thanks{M.T. Askari and L. Lampe are with the Department
of Electrical and Computer Engineering, The University of British Columbia, Vancouver,
BC V6T 1Z4, Canada, e-mail: (mohammadtaha@ece.ubc.ca, lampe@ece.ubc.ca).}
}


\IEEEspecialpapernotice{(Invited Paper)}
\maketitle

\begin{abstract}
Optimizing the input probability distribution of a discrete-time channel is a standard step in the information-theoretic analysis of digital communication systems. Nevertheless, many practical communication systems use transmission based on uniformly and independently distributed symbols drawn from regular constellation sets. The introduction of the probabilistic amplitude shaping architecture has helped to renew interest in using optimized probability distributions, i.e., probabilistic shaping. Traditionally, probabilistic shaping has been employed to reduce the transmit power required for a given information rate over additive noise channels. While this translates into substantive performance gains for optical fiber communication systems, the interaction of shaping and fiber nonlinearity has posed intriguing questions. At first glance, probabilistic shaping seems to exacerbate  nonlinear interference noise (NLIN) due to larger higher-order standardized moments. Therefore, the optimization of shaping distributions must differ from those used for linear channels. Secondly, finite-length effects related to the memory of the nonlinear fiber channel have been observed. This suggests that not only the marginal input-symbol distribution should be looked at. 
In this paper, we provide a tutorial-style discussion of probabilistic shaping for optical fiber communication. Since the distinguishing property of the channel is the signal-dependent NLIN,  we speak of  probabilistic shaping for nonlinearity tolerance. Our analysis builds on the first-order time-domain perturbation approximation of the nonlinear fiber channel and revisits the notion of linear and nonlinear shaping gain. We largely focus on probabilistic amplitude shaping with popular types of shaping methods, {and examine a linear filter model to explain several phenomena associated with probabilistic shaping for fiber nonlinearity, including the interaction between carrier phase recovery and fiber nonlinearity.} The concept of shaping via sequence selection is given special consideration, as it inherently optimizes a  multi-variate distribution for shaped constellations. {We explore how using sign bits for sequence selection offers benefits beyond amplitude shaping, and we introduce a sign-dependent selection metric based on the perturbation model.}

\end{abstract}


\section{Introduction}
\label{s:intro}

Signal shaping in its general form encompasses matching all aspects of a transmit signal to the  properties of the overall communication channel, including  transmitter and receiver constraints. More narrowly, shaping methods have mainly been concerned with reducing average transmit power \cite{fischer:2005}. {Furthermore, since usually pulse-amplitude modulation (PAM) transmission together with constant pulse shapes are considered, shaping often refers to constellation shaping.} 
Constellation shaping permits average-power savings of up to 1.53~dB for a given data rate over the additive white Gaussian noise (AWGN) channel. This so-called shaping gain is approximately additive to the coding gain in coded modulation \cite{Forney:1989,kschischang:1995}.  

Probabilistic constellation shaping, or probabilistic shaping for short, has become a highly popular approach to shaping in optical fiber communication \cite{Cho:2019}. By probabilistic shaping  we mean nonuniform signaling \cite{Forney:1989,kschischang:1995} over the constituent one-dimensional PAM constellation. The relatively recently proposed probabilistic amplitude shaping (PAS) architecture \cite{bocherer2015bandwidth} connects probabilistic shaping in a modular manner with the de-facto standard bit-interleaved forward error-correction (FEC) coded quadrature-amplitude modulation (QAM) transmission. This means that transmitter and receiver processing associated with coding and modulation can remain largely unchanged and that the coding gain from the applied FEC is retained. In addition to this, PAS also enables a fine-grained rate adaptation for a fixed-rate FEC \cite{bocherer2015bandwidth}.


Optimizing probabilistic shaping for minimum average power for the AWGN channel leads to a distribution that chooses signal points with larger amplitudes relatively less frequently.
The resulting power gain will be referred to as \textsl{linear shaping gain}. In the asymptotic case of infinitely many signal points, we obtain a Gaussian distribution and the mentioned 1.53~dB gain. For the optical fiber channel, power saving in itself is not a performance objective, and the maximum signal-to-noise ratio (SNR) is achieved at an optimum launch power. 
The linear shaping gain is important as it permits a higher achievable rate for a given SNR and an extended reach for a required {information rate}.
Furthermore, as increasing the launch power beyond its optimum value is detrimental due to growing nonlinear interference noise (NLIN), another goal of shaping is to enhance nonlinearity tolerance. Accordingly, we also speak of a 
\textsl{nonlinear shaping gain} for optical fiber communication \cite{dar2014shaping}.  


In this paper, we review the interplay of shaping and fiber nonlinearity and discuss the role of probabilistic shaping for improved nonlinearity tolerance. The primary method to gain insights is the time-domain first-order perturbation approximation for the solution of the nonlinear Schr\"odinger equation, which governs the fiber signal propagation \cite{Mecozzi:2012}. It permits us to analyze the impact of the moments of the signal constellation on NLIN and thus the nonlinear shaping gain. Secondly, the memory of the fiber channel due to linear dispersion, which manifests in {similar perturbation coefficients acting on adjacent data terms}, suggests that non-independently and identically distributed (non-i.i.d.) signaling can be favorable \cite{Agrell:2014}. Non-i.i.d.\ signaling applied to shaping can be interpreted as ``shaping out'' NLIN \cite{dar2014shaping}. Interestingly, in the context of PAS, this form of  improved nonlinearity tolerance has been observed as a by-product of practical shaping methods \cite{amari2019introducing, fehenberger2019analysis}.

Our analysis of probabilistic shaping for nonlinearity tolerance is organized as follows. To set the stage, we first briefly revisit signal shaping for Gaussian channels and its connection to the optical fiber channel in Section~\ref{s:ShapingGauss}. 
We then introduce the first-order time-domain perturbation-based model for NLIN and probabilistic shaping optimized for it in Section~\ref{s:ShapingFiber}. We also make the notion of the nonlinear shaping gain concrete. After this, in Section~\ref{s:PAS}, we study the moment and memory effects when applying PAS for shaping, and we present a linear filter model for part of NLIN that manifests as phase noise. This model is shown to be a valuable tool for understanding the nonlinearity tolerance of PAS with practical shaping methods. This includes the mapping from shaped PAM sequences to dual-polarized QAM transmit signals and the presence of a carrier-phase estimation at the receiver. The filter model and related analytical expressions have been used as metrics for PAS with sequence selection. Sequence selection was originally introduced in \cite{secondini2022new} as a means to quantify lower bounds on the capacity of optical fiber channels. In Section~\ref{s:SS}, we discuss the integration of candidate generation and selection metric for sequence selection into PAS. Overall, this results in a probabilistic shaping structure that provides the best nonlinearity tolerance. We conclude with reflections and forward-looking remarks in Section~\ref{s:conclusions}.

This paper is based on a tutorial talk presented at the 2024 Optical Fiber Communication Conference (OFC). Its main objective is to categorize and explain principles that are behind the nonlinearity tolerance achieved with probabilistic shaping. To accomplish this, we organize a large body of literature on the subject matter and also contribute new results, in particular the extended filter model in Section~\ref{s:PAS} and the selection metric in Section~\ref{s:SS}. Most of our quantitative results are for one long-haul fiber link setup, as the purpose is not to explore the best system settings but to illustrate concepts.  We hope that this paper will serve as a helpful entry point for researchers who have not studied shaping for optical fiber channels yet, and as a meaningful point of reference for experts on this topic.


\begin{figure}
\center
\includegraphics[width=0.45\textwidth]{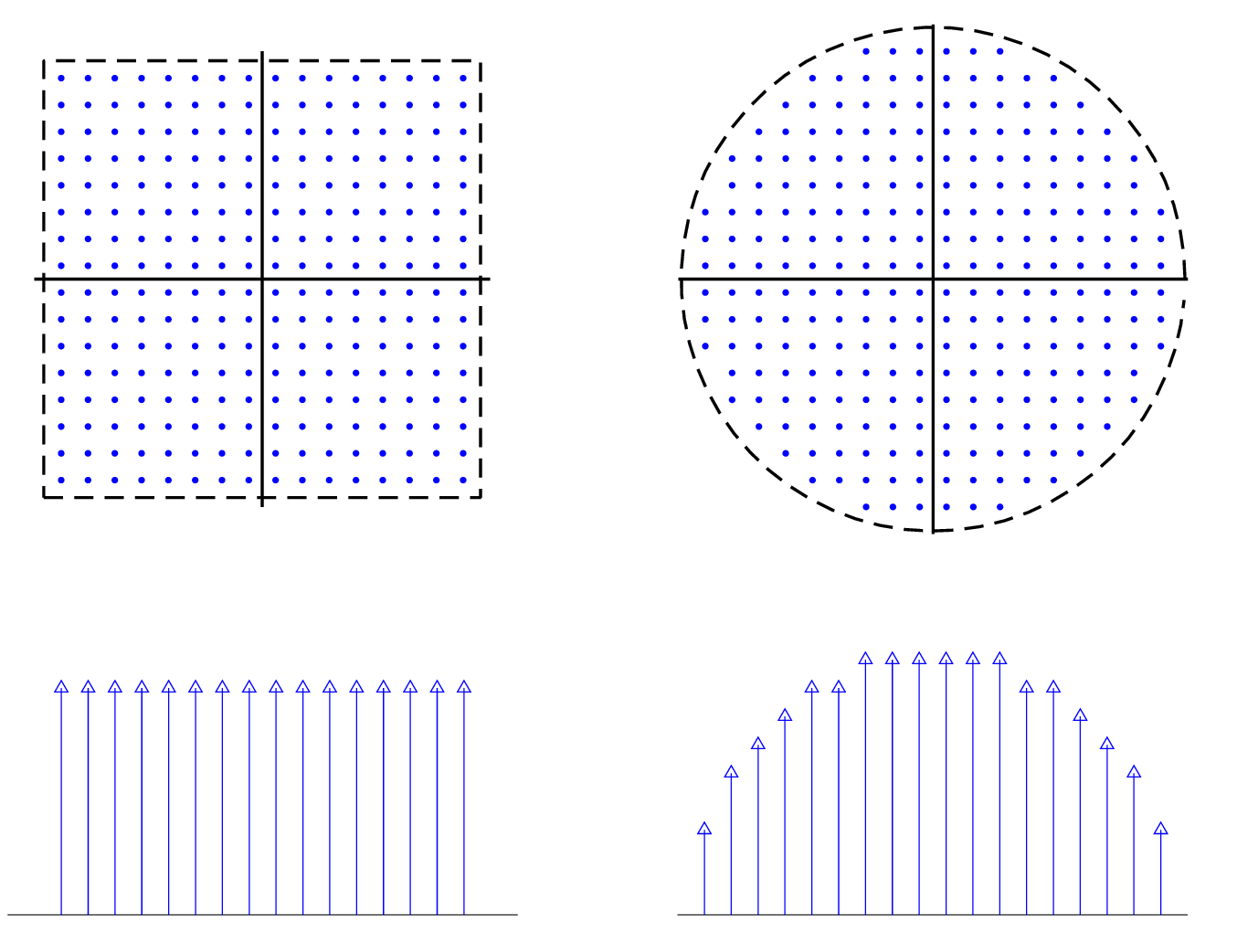}
\caption{Top row: Unshaped (left) and shaped (right) two-dimensional constellation with 256 equi-probable signal points. Bottom row: induced one-dimensional distribution. From \cite[Fig.~4.1]{fischer:2005}.}
\label{f:shaping}
\end{figure}

\section{Signal Shaping for the Gaussian Channel}
\label{s:ShapingGauss}

We start by considering an AWGN channel model and introduce constellation shaping through nonuniform signaling and potential performance gains for optical fiber channels under this model. 

\subsection{Nonuniform Signaling}

We represent signals as points in a $D$-dimensional space, where $D=1$ for i.i.d.\  PAM. Classically, constellation shaping selects $D$-dimensional symbols equi-probably and jointly with $D>1$. 
As mentioned in Section~\ref{s:intro}, a typical objective is to minimize the variance of the signal constellation  for a given information rate. 
Borrowing the example from \cite[Fig.~4.1]{fischer:2005}, the top row of Figure~\ref{f:shaping} shows an instance of unshaped (left) and shaped constellation (right) for $D=2$. The bottom row of Figure~\ref{f:shaping} presents the probability distribution for the underlying one-dimensional constellation that is induced by shaping. This suggests an alternative view of shaping, which is the use of  nonuniform signaling, or probabilistic constellation shaping \cite{kschischang:1995,fischer:2005}. This can also directly be applied to $(D=1)$-dimensional PAM. 

\begin{figure}
\center
\includegraphics[width=0.45\textwidth]{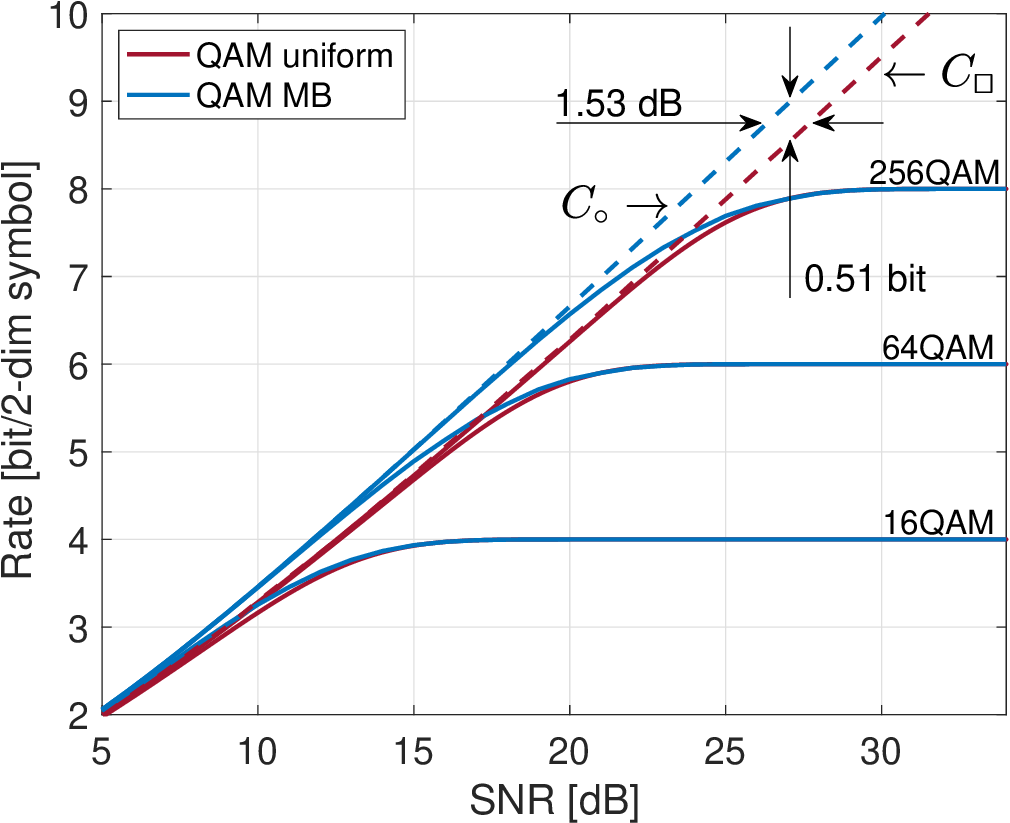}
\caption{Achievable rates per two dimensions for different uniform and MB-shaped constellations and the AWGN channel. Dashed lines: $C_\circ$ and $C_{\mbox{\tiny$\square$}}$ are the rates for continuous Gaussian and uniform signaling, respectively. Solid lines: QAM uniform and QAM with MB shaping. }
\label{f:capacityAWGN}
\end{figure}

The Maxwell-Boltzmann (MB) distribution takes a special place among the probability distributions for nonuniform signaling. For a $D$-dimensional constellation ${\cal A}=\{x_1,\ldots,x_\CS\}$ with probability distribution ${\cal P}=\{p_1,\ldots,p_\CS\}$, the MB distribution is given by
\begin{equation}
p_i=\e^{-\lambda \|x_i\|^2} / \sum\limits_{x\in{\cal A}}\e^{-\lambda\|x\|^2}, 
\end{equation}
with parameter $\lambda \ge 0$ to adjust the trade-off between entropy, i.e., rate, and variance of $({\cal A, P})$. As discussed in \cite{kschischang:1995}, \cite[Ch.~4]{fischer:2005}, the MB distribution has several optimality properties. For example, as it is illustrated in Figure~\ref{f:capacityAWGN} for the case of QAM transmission, MB-shaping permits approaching the ultimate shaping gain of 1.53~dB in SNR for the AWGN channel for any $D$. For this reason, the MB constellation has also widely been used for probabilistic shaping in optical fiber communications.

\subsection{Gaussian Noise Model for the Nonlinear Fiber Channel}

Since propagation through the fiber channel is governed by the nonlinear Schr\"odinger equation, which does not permit an analytical solution, simplified phenomenological models for the input-output relationship of fiber signals have been developed. An important such model is the Gaussian noise (GN) model \cite{Poggiolini:2014}. It accounts for the NLIN as AWGN with  power $\Pnlin=\eta P^3$, where $P$ is the transmit power per wavelength division multiplexing (WDM) subchannel\footnote{For simplicity, we assume that the same transmit power $P$ is used in all WDM channels.}, and $\eta$ is independent of $P$. Accordingly, the effective SNR is expressed as  
\begin{eqnarray}
\SNRnl&=&\frac{P}{\pase+\Pnlin}=\frac{P}{\pase+\eta\cdot P^3},
\label{e:SNReff}
\end{eqnarray}
where $\pase$ denotes the accumulated amplifier spontaneous emission (ASE) noise. Approximating the contributions to NLIN from different spans as additive as per the incoherent GN (IGN) model, we have \cite{Poggiolini:2014}
\begin{equation}
\SNRnl=\frac{P}{(P_{\mathrm{ASE,1}}+\eta_1 P^3)N_{\mathrm{s}}}\le \frac{1}{3N_{\mathrm{s}}}
\left(\frac{4}{P_{\mathrm{ASE,1}}^2\eta_1}\right)^{1/3},
\label{e:SNRIGN}
\end{equation}
where $P_{\mathrm{ASE,1}}$ and $\eta_1P^3$ are the ASE from  a single amplifier and the NLIN generated 
in a single span, respectively. The inequality follows from operating at the SNR-optimum transmit power 
\begin{equation}
\label{e:PoptSNR}
\Popt=\left(\frac{\pase}{2\eta}\right)^{1/3}\;.
\end{equation}

\begin{figure}
\center
\includegraphics[width=0.45\textwidth]{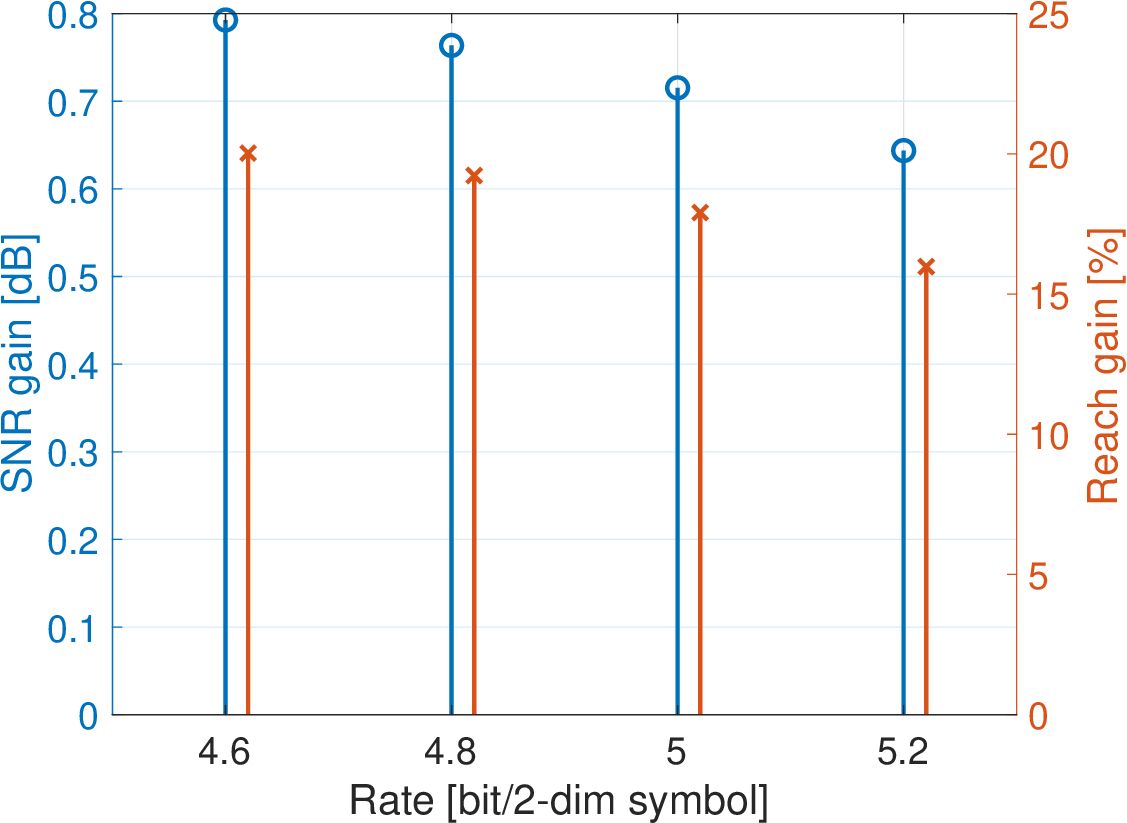}
\caption{SNR gain (left) and reach gain (right) for MB-shaping of a 64QAM constellation for several achievable rates. IGN model at optimum transmit power.}
\label{f:GNrange}
\end{figure}

Inequality \eqref{e:SNRIGN} permits us to relate SNR gains from shaping as shown in Figure~\ref{f:capacityAWGN} to gains in maximum reach. This is illustrated in Figure~\ref{f:GNrange} for a 64QAM constellation and several meaningful target rates. We observe a reach extension of 15\% to 20\%, which substantiates the significant interest in shaping for the nonlinear fiber channel.

\section{Signal Shaping for {the} Nonlinear Fiber Channel}
\label{s:ShapingFiber}

The GN model for the optical fiber channel allows us to quantify potential benefits from shaping associated with the linear shaping gain. However, it fails to capture the effect of shaping on nonlinearity tolerance, which in part manifests in amplitude modulation induced noise (AMIN).  
\subsection{Amplitude Modulation Induced Noise}


The GN model is derived from a perturbation assumption for a sufficiently small nonlinearity together with a Gaussianity assumption for the transmitted signal. The more refined enhanced GN (EGN) model\cite{Carena:2014} dispenses with the second assumption. 
We follow this approach to identify the AMIN \cite{dar2014shaping}.

Using the time-domain first-order perturbation approximation, 
we can express the NLIN after linearity compensation and sampling as \cite{Mecozzi:2012,Dar:2013}
\begin{equation}
\label{e:perturbation}
\Delta x_{k,0}=\sum\limits_{m,n,p,s} x_{m+k,0}x_{n+k,s}x^*_{p+k,s}\chi_{m,n,p,s}\;.
\end{equation}
In~\eqref{e:perturbation}, we consider the NLIN for the $k$-th data symbol in the channel of interest $s=0$,  $x_{k,s}$ denote the data from WDM channel $s$, and $\chi_{m,n,p,s}$ are perturbation coefficients for intra-channel ($s=0$) and inter-channel ($s\neq 0$) NLIN. Computing the variance $\mathbb{E}(|\Delta x_k|^2)$ of  $\Delta x_k$ reveals that the intensity of the NLIN is still proportional to $P^3$, {but the coefficient $\eta$ differs from that for the GN model by an additive term which is a function of fourth ($\mu_4$) and sixth ($\mu_6$) standardized moment of the constellation used for transmission. Exactly this is the effect of AMIN. It also means that the constellation interacts with NLIN and leads to the notion of a nonlinear shaping gain.\footnote{We note that the GN and EGN models have recently been  refined to account for stimulated Raman scattering, which becomes significant for ultra-wideband transmission systems \cite{semrau2018gaussian, Poggiolini:2018, semrau2019modulation}.}}
%

\begin{figure}
\includegraphics[width=0.45\textwidth]{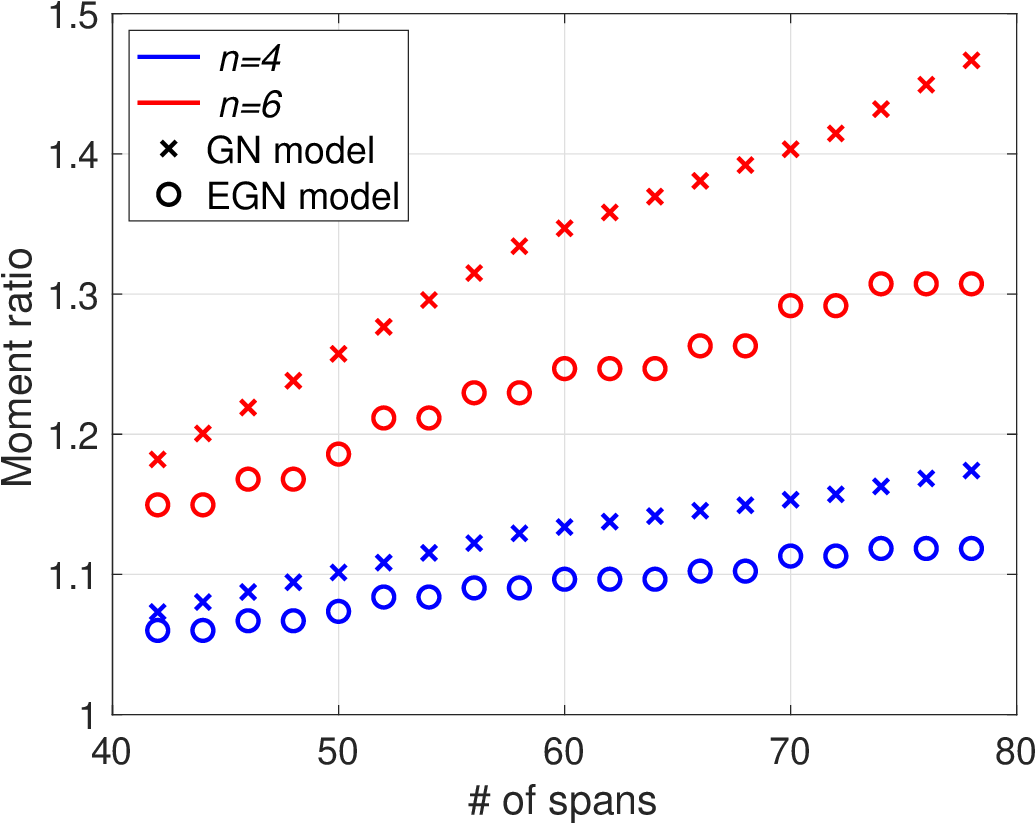}
\caption{Ratio $\mu_n^{\mathrm{sh}}/\mu_n^{\mathrm{uni}}$, $n=4,6$, of the standardized moments for shaped and uniform 16QAM constellations, where the shaped constellations were obtained in \cite{pan2016probabilistic} from optimizing the mutual information for fiber links of length (\# of spans)$\times80$~km. The EGN model was used for the optimization. Also shown are the moment ratios for the shaped constellation that would have been obtained with the GN model.}
\label{f:momentratio}
\end{figure}
The study in \cite{pan2016probabilistic}  used the EGN model to optimize the probability distribution for nonuniform 16QAM transmission in a long-haul fiber system. 
Simulation results showed a more than 10\% reach increase due to shaping.\footnote{While the amplitude distribution was optimized directly in \cite{pan2016probabilistic}, practically identical results were obtained for optimizing an MB distribution.} This is somewhat less than what would have been expected from the GN model (as in Figure~\ref{f:GNrange}). The reason for this is that nonuniform signaling increases AMIN compared to uniform QAM. 

To illustrate this, Figure~\ref{f:momentratio} shows the ratio of standardized moments $\munsh$ for shaped and $\mununi$ for uniform 16QAM and $n=4,6$, as a function of the number of $80$~km spans in the link. The curve labeled ``EGN model'' is for the optimized shaped constellations obtained in \cite{pan2016probabilistic}. The curve labeled ``GN model'' corresponds to the moment ratios that would have been obtained assuming a GN model. We observe that (i) all ratios are greater than one, i.e., AMIN increases due to shaping, and (ii) the EGN model, which accounts for AMIN, forces a reduction of these moments. In other words, the EGN model also considers the nonlinear shaping gain.


\subsection{Nonlinear Shaping Gain}

The notion of the nonlinear shaping gain has been made concrete in \cite{dar2014shaping,Geller:2016}. Following \cite{Geller:2016}, we consider the 
so-called system quality factor (SQF)
\begin{equation}
\label{e:sqf}
Q = \frac{\dminsq}{\pase+\Pnlin},
\end{equation}
where $\dminsq$ is the squared minimal Euclidean distance for a signal constellation. The SQF is directly related to the error rate performance for transmission over the EGN channel. As such, it takes a role  analogous to the so-called constellation figure of merit (CFM) $\dminsq/P$ used for shaping over the AWGN channel  \cite{kschischang:1995}, \cite[Ch.~4]{fischer:2005}.  
We can develop \eqref{e:sqf} as
\begin{equation}
\label{e:qmax}
Q \stackrel{(a)}{=}\frac{\nu P}{\pase+\eta P^3}\stackrel{(b)}{\le}\frac{1}{3}\left(\frac{\pase}{2}\right)^{-\frac{2}{3}} \frac{\nu}{\eta^{1/3}}\eqqcolon\Qmax,
\end{equation}
where (a) we used that $\dminsq$ is proportional to the transmit power and that NLIN is proportional $P^3$ given the perturbation model \eqref{e:perturbation}, and (b) we considered the optimum transmit power from \eqref{e:PoptSNR}.
 Both $\nu$ and $\eta$ are functions of the used signal constellation ${\cal A}$ and the applied probabilistic shaping ${\cal P}$. We can now measure the shaping gain with respect to a reference system when operating at optimum power as \cite{Geller:2016} 
\begin{equation}
\label{e:gainlinnlin}
\frac{\Qmaxsh}{\Qmaxref} =   \frac{\nush}{\nuref}\left(\frac{\etaref}{\etash}\right)^{\frac{1}{3}}\;.
\end{equation}

Expression \eqref{e:gainlinnlin} reveals a partitioning into a linear shaping gain $\glin={\nush}/{\nuref}$ and a nonlinear shaping gain $\gnlin=({\etaref}/{\etash})^{{1}/{3}}$. The former is the gain known from the AWGN channel and maximized with the MB distribution for a given rate. The latter is related to AMIN. It can equivalently be expressed in terms of the differences of maximum effective SNR \eqref{e:SNRIGN} or optimum transmit power  \eqref{e:PoptSNR} for shaped and unshaped transmission, {i.e.,}
{
\begin{equation}
\label{e:gainlin_snr_power}
\frac{\SNRmaxsh}{\SNRmaxref} = \frac{\Poptsh}{\Poptref} = \left(\frac{\etaref}{\etash}\right)^{\frac{1}{3}} = \gnlin\;.
\end{equation}
}

Comparing i.i.d.\ uniform and MB-shaped QAM signaling over an optical link, the linear shaping gain $\glin$ in dB is positive, while the nonlinear shaping gain $\gnlin$ in dB is negative due to the increased standardized fourth and sixth moments, i.e., higher AMIN. 

It has been pointed out in \cite{dar2014shaping} that the nonlinear shaping gain can be significantly positive for sphere shaping with $D>1$. In fact, the aggregate shaping gain \eqref{e:gainlinnlin} can exceed the $1.53$~dB ultimate shaping gain for the AWGN channel. The underlying reason is the channel memory due to linear dispersion that manifests in the variation of perturbation coefficients $\chi_{m,n,p,s}$ in \eqref{e:perturbation} with time lags $m,n,p$. Then, using sphere shaping with a $D$-dimensional constellation such that $D$ is related to the channel memory \cite{Agrell:2014} reduces the variance of AMIN \cite{dar2014shaping,Geller:2016}. 
This effect is not captured by only considering standardized moments. We will elaborate on this 
in the context of PAS in the following.

\section{Probabilistic Amplitude Shaping}
\label{s:PAS}

In this section, we consider the PAS coded modulation from  \cite{bocherer2015bandwidth,Boecherer:2019} as the underlying architecture to study shaping for nonlinearity tolerance. The advantage of PAS is the decoupling of FEC and constellation shaping, where the latter is realized through an amplitude shaper. In the following, we will focus on specific implementations of the amplitude shaper and discuss the interplay of linear and nonlinear shaping gains in connection to standardized moments and channel memory. 


\subsection{Shaping Methods}

The amplitude shaper in PAS introduces a nonuniform signaling for the amplitude levels of a PAM constellation. In the initial work \cite{bocherer2015bandwidth}, this is done by a distribution matcher  using constant composition distribution matching (CCDM) \cite{schulte2015constant}, typically to approach an MB distribution. An alternative method introduced in \cite{amari2019introducing} uses sphere shaping in the form of enumerative sphere shaping (ESS) \cite{Willems:93}. Hence, CCDM and ESS are representatives of two types of shaping methods that directly target and indirectly induce probabilistic shaping for PAM signaling, respectively \cite{gultekin2020probabilistic}. Even though CCDM starts with a target distribution and ESS does not, they have in common that they operate on blocks of symbols. Accordingly, we denote the block length by $D$. 
 While ESS permits blocks of PAM symbols that lie within a $D$-dimensional sphere, CCDM chooses PAM blocks that are permuted versions of each other, i.e., they can be represented as being located on a sphere in $D$ dimensions. 

\begin{figure}[t]
\includegraphics[width=0.45\textwidth]{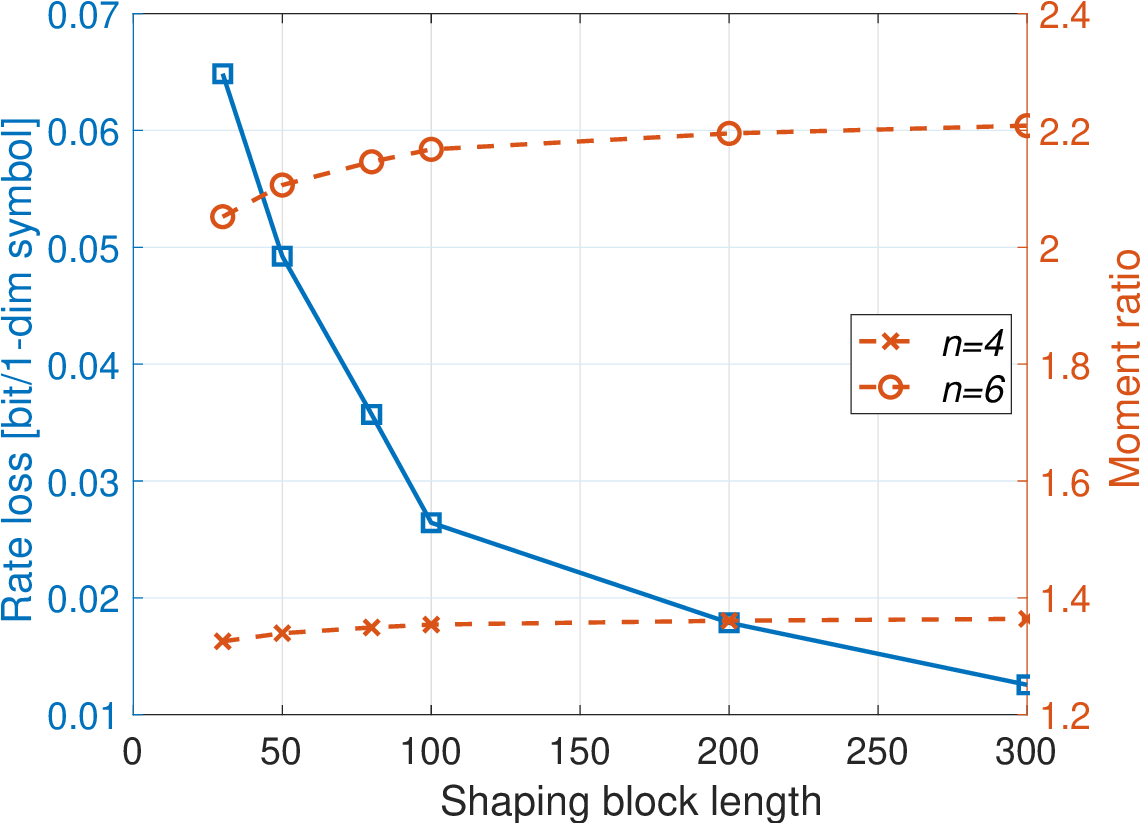}
\caption{Rate loss (left) and {ratio} $\munsh/\mununi$, $n=4,6$ (right) for ESS with 8PAM and $\Rshaper=2.5$~bits per amplitude versus shaping block length. Standardized moments are for 64QAM obtained from pairing two 8PAM  ESS output blocks. }
\label{f:rateloss}
\end{figure}
Besides differences in computational complexity and incurred transmission latency for these and other shaping methods for finite block length $D$ (see e.g.\ \cite{Fehenberger2020huffman}), a key performance criterion is the rate loss
\begin{equation}
\label{e:rloss}
\Rloss = H({\cal A},{\cal P})-\Rshaper\;.
\end{equation}
  In \eqref{e:rloss}, $\Rshaper$ denotes the rate of the shaping method in bits per PAM symbol and  $H({\cal A},{\cal P})$ is the entropy of the resulting shaped PAM constellation. ESS generally provides a lower rate loss than CCDM for short block lengths $D$. 
  
  Figure~\ref{f:rateloss} (left $y$-axis) shows the rate loss for ESS and 8PAM with $\Rshaper=2.5$~bit/1-dimension as a function of block length $D$.  Mapping length-$D$ 8PAM output sequences to each one of the quadrature components of a 64QAM constellation, the figure (right $y$-axis) also shows the ratios $\munsh/\mununi$ for the $n=\{4,6\}$-th standardized moments, which are relevant for AMIN. We observe that rate loss decreases with block length and reaches zero for $D\to\infty$ when the ESS approaches the MB distribution \cite{Gultekin:2018}. On the other hand, since the moment ratios are notably larger than one and slightly increase with $D$, we expect AMIN to grow because of PAS. 

\subsection{Moment Effects}
\label{e:moments}

\subsubsection{Demonstration} 
\begin{figure}[t]
\includegraphics[width=0.45\textwidth]{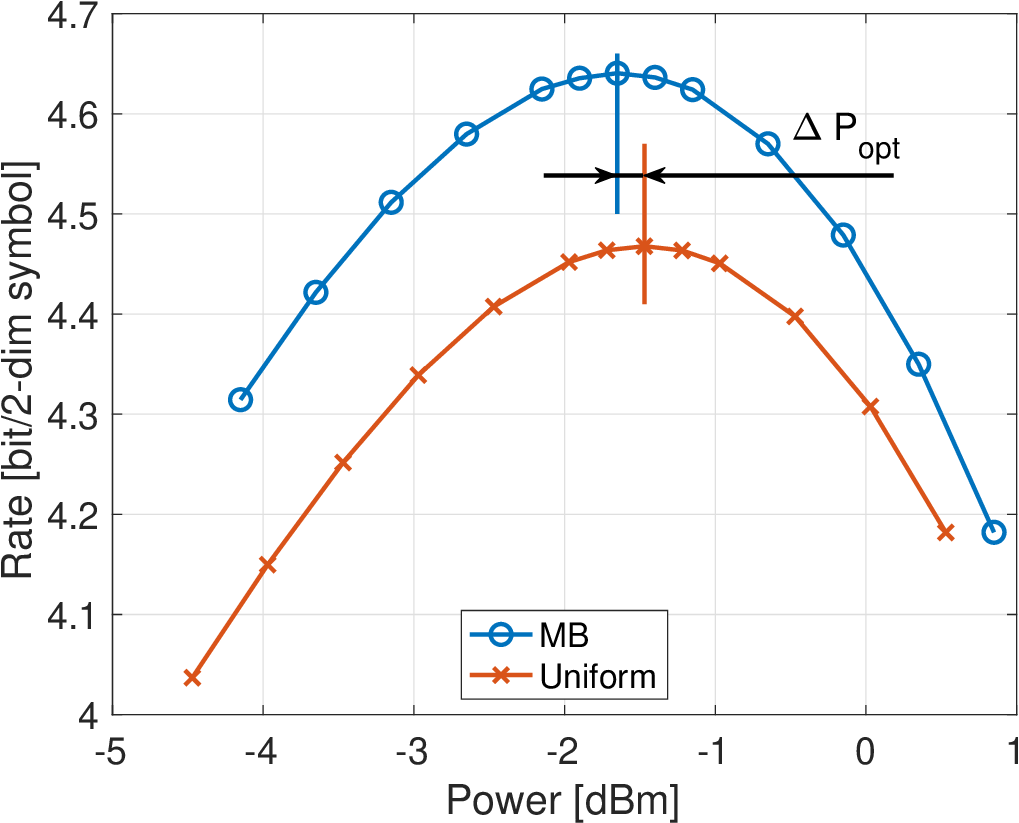}
\caption{Achievable rate per two dimensions for 64QAM WDM transmission over a 2000~km link (system settings from \cite[Table~III]{fehenberger2016probabilistic}). PAS with optimized MB and uniform distribution.}
\label{f:SNRloss}
\end{figure}

The effect of PAS on AMIN has indeed been observed and studied in \cite{fehenberger2016probabilistic, fehenberger2019analysis}. We illustrate it based on the optical fiber system example from \cite{fehenberger2016probabilistic}, which considers a 64QAM transmission at 26~Gbaud and 9~WDM channels over 20 spans of 100~km (see \cite[Table~III] {fehenberger2016probabilistic} for further details). We use the EGN model for the following results, as they have been found to closely match results from simulations with the split-step Fourier method (SSFM) in \cite{fehenberger2016probabilistic}.  
Figure~\ref{f:SNRloss} shows the achievable rate obtained from the rate-SNR relations in Figure~\ref{f:capacityAWGN} as a function of transmit power per WDM channel when using uniform and shaped 64QAM constellations. For the latter, we apply MB shaping with $\lambda=0.03$, {which is the result of a line search to maximize rate at optimum power}. MB shaping corresponds to PAS with CCDM or ESS and a sufficiently large block length $D$ so that $\Rloss\to 0$. As can be seen from the figure, PAS yields a notable rate gain. The optimum transmit power for shaped transmission is reduced by $\Delta\Popt=0.18$~dB compared to uniform QAM. This means that the negative nonlinear shaping gain $\gnlin=-0.18$~dB is experienced in this setting. The gain is thus a loss, which is not a contradiction as the total shaping gain is what is being optimized. The loss would have been higher if $\lambda$ had been chosen according to the GN model and the effective SNR achieved by uniform transmission. Hence, the shaping design for a reduced optimal transmit power is a form of enhanced nonlinearity tolerance. The linear shaping gain for MB shaping with the selected $\lambda$ is about $\glin=(2^{2H(\cal{A},{\cal P})}-1)\dminsq/(6P)=1.18$~dB.


\subsubsection{Nonlinearity tolerance}

Attempts to increase 
the nonlinear shaping gain can broadly be divided into two categories. 

The first is to explicitly design a suitable shaping distribution. This could be an MB distribution with a shaping parameter $\lambda$ that does not maximize the linear shaping gain, as in the example in Figure~\ref{f:SNRloss}.
Or it could be a distribution that is directly optimized for the nonlinear fiber channel and different from the MB distribution. An instance of this is the result in Figure~\ref{f:momentratio} for the distribution design from \cite{pan2016probabilistic}. To obtain the distributions, EGN or EGN-type models and grid search \cite{pan2016probabilistic} or Blahut-Arimoto algorithm-like iterative search \cite{Yankov:2016} have been used. The EGN model also inspired a so-called super-Gaussian distribution design in \cite{Tehrani:2018}. Other examples are based on the {approximation} 
\begin{equation}
\label{e:EGNkurtosis}
\etash \approx \eta_1+\eta_2(\mufsh-2),
\end{equation}
{for the EGN-model coefficient, where $\eta_1,\eta_2>0$ depend on the fiber link parameters \cite{Dar:2013,Carena:2014}}, noting that excess kurtosis dominates AMIN. For example, \cite{Sillekens:2018} introduces a two-parameter distribution that includes the second and the fourth moment. 

\begin{figure}[t]
\includegraphics[width=0.48\textwidth]{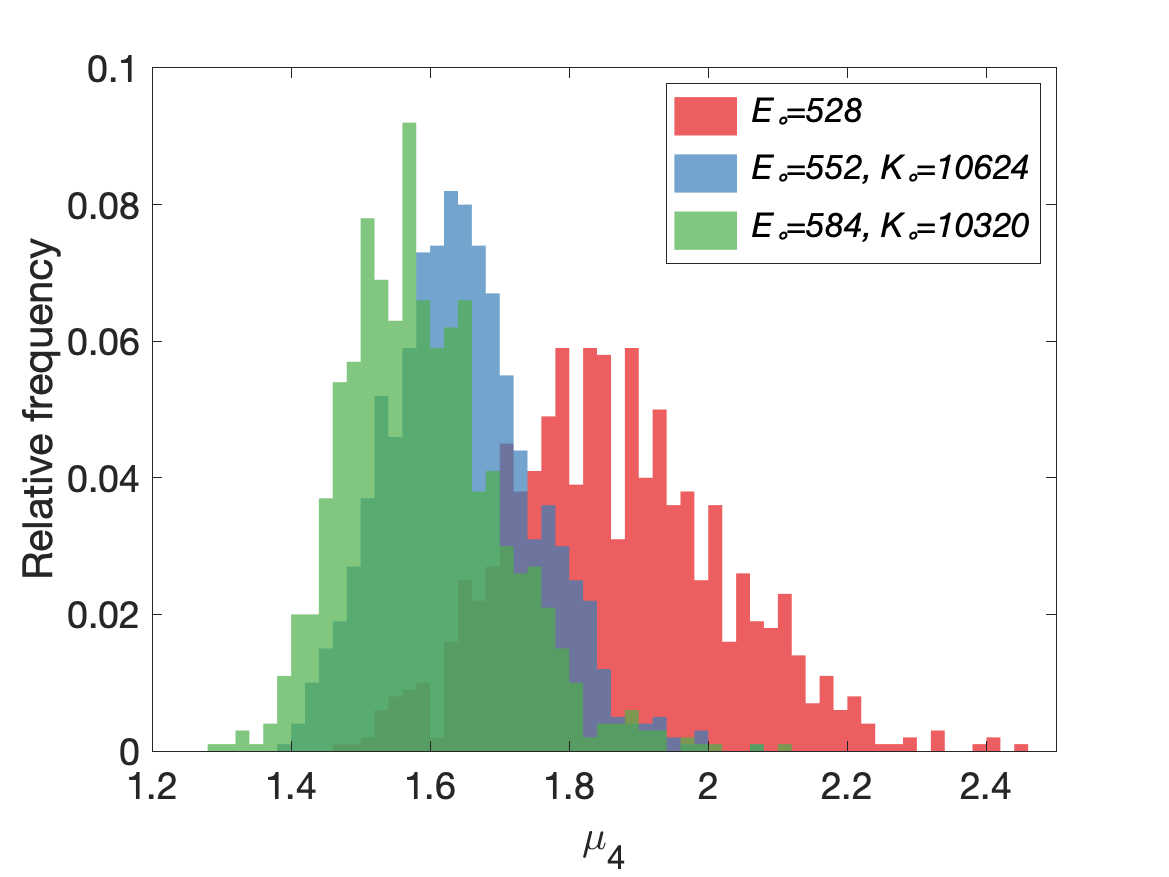}
\caption{Empirical distribution of the kurtosis of 64QAM symbols obtained from K-ESS for 8PAM with block length $D=64$ and shaping rate $2.5$~bit/1-dim symbol. $E_\circ$ and $K_\circ$ are the constraints on $\sum_{i=1}^{D}a_i^2$ and $\sum_{i=1}^{D}a_i^4$, respectively. $E_\circ=528$ corresponds to ESS. Example from \cite[Fig.~3]{gultekin2021kurtosis}.}
\label{f:KESS}
\end{figure}

The second category modifies sphere shaping or distribution matching (DM) algorithms of PAS to mitigate nonlinearity. A prominent example of the former is kurtosis-limited ESS (K-ESS) 
\cite{gultekin2021kurtosis}. Following \eqref{e:EGNkurtosis}, it extends ESS to apply a constraint on empirical fourth moments of amplitude sequences in addition to their energies. Figure~\ref{f:KESS} shows the empirical distributions of the kurtosis of 64QAM symbols obtained from K-ESS for 8PAM with a block length of $D=64$ and a shaping rate of $2.5$~bit/1-dimension The different histograms correspond to different  ($E_\circ,K_\circ)$ pairs for the constraints on amplitude-sequence energies, i.e., $\sum_{i=1}^{D}a_i^2\le E_\circ$, and fourth-moment sums, i.e.,  $\sum_{i=1}^{D}a_i^4\le K_\circ$. We observe a trade-off between energy (equivalent to second moment) and kurtosis, which in turn permits to balance linear and nonlinear shaping gains. Another example that aims at reducing excess kurtosis is the hierarchical distribution matching (HiDM) in \cite{Yoshida:2020}, which sorts candidate sequences based on their $(n/2)$-th moment and searches over $n=1,2,\ldots,8$.

\subsection{Finite-length Effects}

The moment effects and mitigation thereof described above are a direct consequence of the EGN model and the notion of AMIN. However, since shaping methods (CCDM, ESS, and alike) are preferably used with shorter shaping block lengths to reduce latency and possibly complexity, finite block-length effects have been observed for PAS over the nonlinear fiber channel.

\subsubsection{Phenomenological observation}
\begin{table}[t]
\caption{System parameters used for numerical results  \cite{wu2022list}}
\label{tab:setup1_sim_param}
\centering
    \begin{tabular}{c | c}     
     \hline\hline
     Parameter & Value \\
     \hline\hline
     Modulation & 256~QAM \\
     Amplitude shaper & CCDM \\
     Shaping rate & 2.4~bits/amplitude \\
     Polarization & Single \\
     Center wavelength  & 1550~nm \\
     Symbol rate & 32~GBd \\ [1ex] 
     WDM spacing & 50~GHz \\ 
     \# WDM channels & 11 \\ 
     { Total bandwidth} & { 550~GHz} \\
     Pulse shape & Root-raised cosine\\ 
     Pulse roll-off & 0.1 \\  
     \hline
     Span length & 80~km\\
     \# Spans & 20 \\
     Fiber loss & 0.2~dB/km \\
     Dispersion parameter & 17~ps/nm/km \\
     Nonlinearity parameter & 1.37~1/W/km \\
     EDFA noise figure & 6~dB \\
     \hline\hline 
    \end{tabular}
\end{table}

\begin{figure}[t]
\includegraphics[width=0.45\textwidth]
{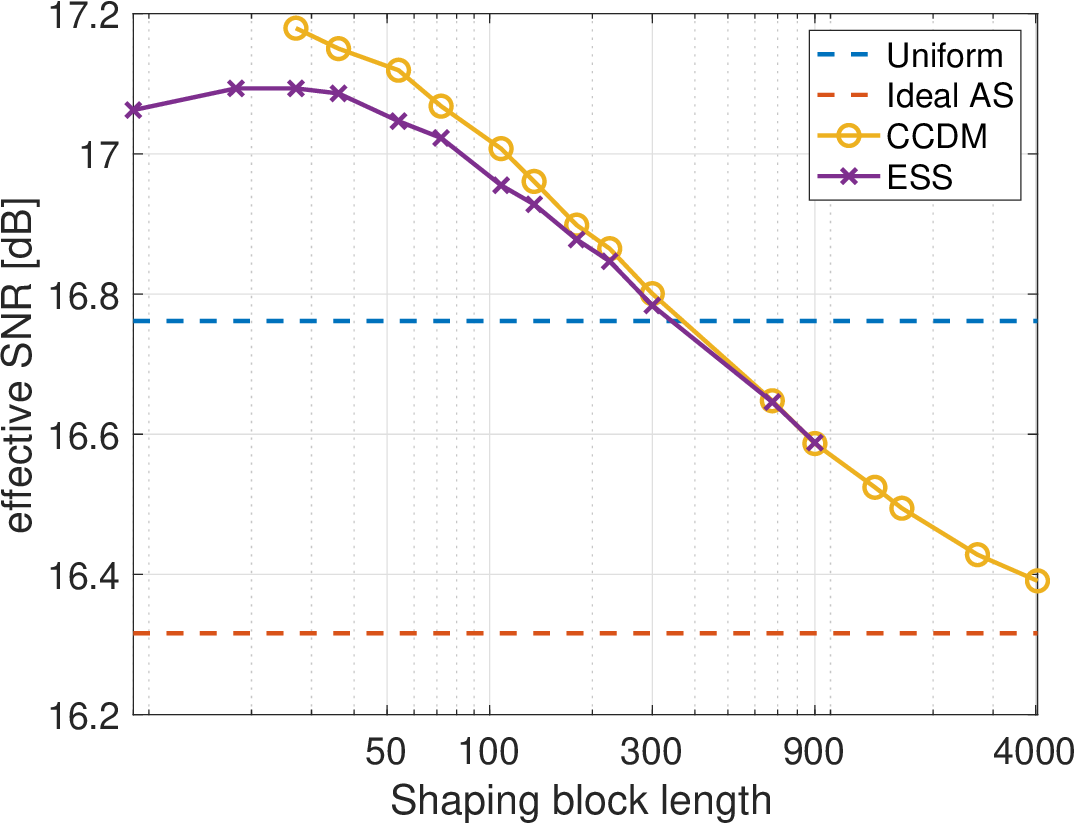}
\caption{Effective SNR versus shaping block length. SSFM-based simulation results for system parameters in Table~\ref{tab:setup1_sim_param}. PAS with CCDM and ESS. Dashed lines are reference curves for shaping with infinite block length (``Ideal AS'') and for transmission with uniform probabilities for signal points (``Uniform''). All results are at optimal transmit power. }
\label{f:SNReffective}
\end{figure}
In particular, it has been observed that the effective SNR increases with shorter shaping block length for both ESS and CCDM \cite{amari2019introducing, fehenberger2019analysis}. This is illustrated  in Figure~\ref{f:SNReffective}, which shows SSFM simulation results for the effective SNR versus shaping block length for PAS with CCDM and ESS. As reference points, the SNRs for ``ideal shaping'' with infinite block length, i.e.,  transmission with symbols drawn i.i.d.\ from an MB distribution, and for transmission with uniform probabilities are included. All results are obtained at an optimal transmit power. The link and system parameters are taken from \cite{wu2022list} and summarized in Table~\ref{tab:setup1_sim_param}. For consistency, we will use the same system parameters with slight modifications throughout the remainder of this paper. 

{We note from Figure~\ref{f:SNReffective} that the effective SNR is a decreasing function of shaping block length, except for ESS with extremely short block lengths,  which will be justified in Section~\ref{s:linearfiltermodel}. In general, shaped transmission with ESS and CCDM and short block lengths provides a higher effective SNR than uniform 256QAM. As the block length increases, the effective SNR approaches the curve of ideal shaping of 256QAM with an MB distribution, which is lower than the SNR of uniform transmission.}

\begin{figure}[t] 
\includegraphics[width=0.45\textwidth]
{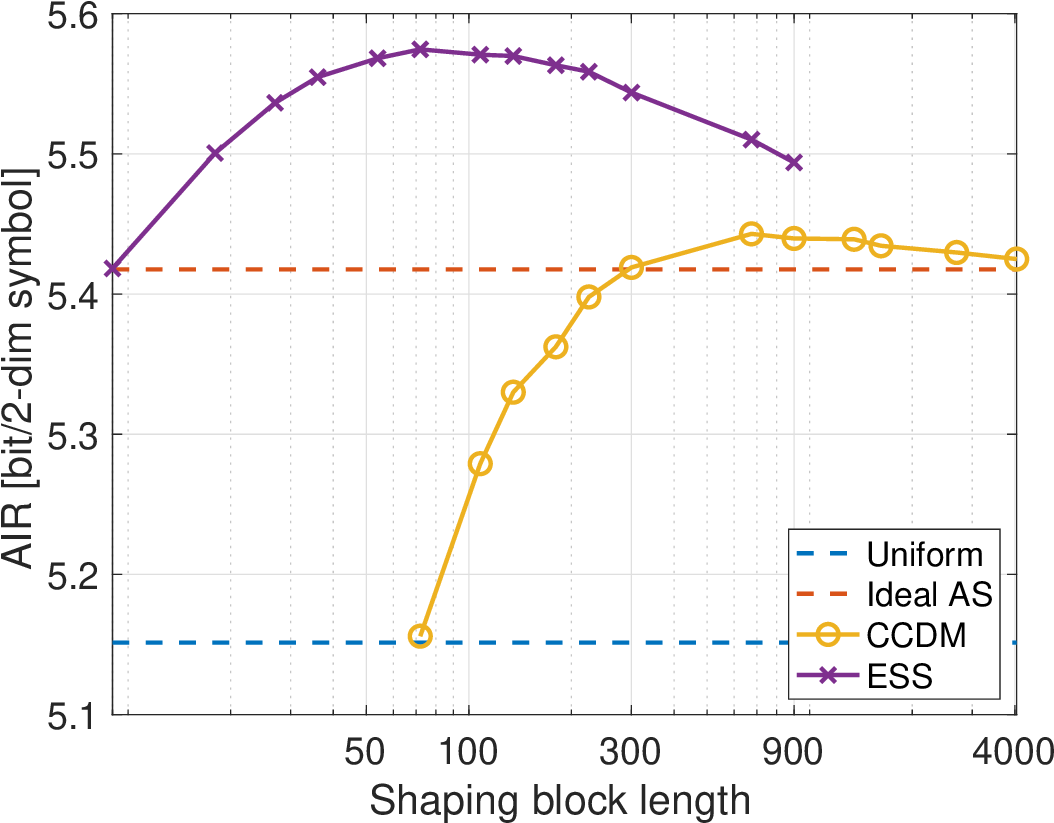}
\caption{AIR per two dimensions versus shaping block length. SSFM-based simulation results for system parameters in Table~\ref{tab:setup1_sim_param}.  PAS with CCDM and ESS. Dashed lines are reference curves for shaping with infinite block length (``Ideal AS'') and for transmission with uniform probabilities for signal points (``Uniform''). All results are at optimal transmit power. }
\label{f:AIRvsBL}
\end{figure}

From \eqref{e:SNReff} and \eqref{e:PoptSNR}, we write the effective SNR at optimum power as 
\begin{equation}
\label{e:SNReffopt}
\SNRnlopt = \frac{1}{3}\left(\frac{2}{\pase}\right)^{\frac{2}{3}} \eta^{-1/3}
\end{equation}
and recall from \eqref{e:qmax} that it is a measure for the nonlinear shaping gain. Hence, we conclude that short block-length shaping improves nonlinearity tolerance.  Since on the other hand rate loss $\Rloss$ in \eqref{e:rloss} increases with decreasing block length, an optimal block length for shaped transmission over the optical fiber channel emerges. This is demonstrated in Figure~\ref{f:AIRvsBL}, which shows the achievable information rate (AIR) as a function of shaping block length. We calculate AIR utilizing the bit-metric decoding method as described in \cite[Eq.~(8)]{fehenberger2016probabilistic}.

\subsubsection{Shaping methods}

The moment effects discussed in Section~\ref{e:moments} are directly understood from computing the second moment, i.e., $\Pnlin$, from the first-order perturbation approximation in \eqref{e:perturbation}. This expression of NLIN also provides the key to appreciating the finite-length effects illustrated in Figure~\ref{f:SNReffective}. In particular, it has been argued in \cite{dar2014shaping} that the perturbation coefficients $\chi_{m,n,p,s}$ vary slowly with delay-time indices\footnote{In \cite{dar2014shaping}, the special case of $\chi_{0,n,n,s}$ accounting for the dominating terms of cross-phase modulation (XPM) was considered.} and thus reducing signal-amplitude fluctuations between blocks of symbols is favorable. This amounts to shaping on a $D$-dimensional sphere, and the results in \cite{dar2014shaping, Geller:2016} identify an optimal shaping block length $D$ for a given nonlinear fiber channel.
Following the same line of argument, finite-state machine sources are used in \cite{Yankov:2017} to introduce dependencies between shaped amplitudes and to render transmission nonlinearity tolerant. 

In the context of PAS, variants of CCDM and ESS have been derived to account for temporal channel effects. For CCDM, the redundancy in the bisection-based CCDM implementation is used to limit energy variations between hyper-symbols consisting of multiple amplitudes \cite{fu2021parallel}. For ESS, a band-trellis version in \cite{gultekin2022mitigating} removes shaped sequences from inner shells and only uses sequences from outer shells with similar radii or energy.







\subsubsection{Analysis}
\begin{figure}[t]
\centerline{%
\begin{tikzpicture}[thick,scale=0.9, every node/.style={transform shape}]
\filldraw[thick, align=center, fill={rgb:orange,1;yellow,2;pink,5}] (0,0) rectangle node{} (0.2,0.4);
\filldraw[thick, align=center, fill={rgb:orange,1;yellow,2;pink,5}] (0.2,0.0) rectangle node{} (0.4,0.4);
\filldraw[thick, align=center, fill={rgb:orange,1;yellow,2;pink,5}] (0.4,0.0) rectangle node{} (0.6,0.4);
\filldraw[thick, align=center, fill={rgb:orange,1;yellow,2;pink,5}] (0.6,0.0) rectangle node{} (0.8,0.4);
\filldraw[thick, align=center, fill={rgb:white,5;pink,2}] (0.8,0.0) rectangle node{} (1,0.4);
\filldraw[thick, align=center, fill={rgb:white,5;pink,2}] (1,0.0) rectangle node{} (1.2,0.4);
\filldraw[thick, align=center, fill={rgb:white,5;pink,2}] (1.2,0.0) rectangle node{} (1.4,0.4);
\filldraw[thick, align=center, fill={rgb:orange,1;yellow,2;blue,3}] (1.4,0.0) rectangle node{} (1.6,0.4);
\filldraw[thick, align=center, fill={rgb:orange,1;yellow,2;blue,3}] (1.6,0.0) rectangle node{} (1.8,0.4);
\filldraw[thick, align=center, fill={rgb:orange,0;yellow,1;blue,2}] (1.8,0.0) rectangle node{} (2.0,0.4);
\filldraw[thick, align=center, fill={rgb:orange,1;yellow,2;pink,5}] (2.0,0) rectangle node{} (2.2,0.4);
\filldraw[thick, align=center, fill={rgb:orange,1;yellow,2;pink,5}] (2.2,0.0) rectangle node{} (2.4,0.4);
\filldraw[thick, align=center, fill={rgb:orange,1;yellow,2;pink,5}] (2.4,0.0) rectangle node{} (2.6,0.4);
\filldraw[thick, align=center, fill={rgb:orange,1;yellow,2;pink,5}] (2.6,0.0) rectangle node{} (2.8,0.4);
\filldraw[thick, align=center, fill={rgb:white,5;pink,2}] (2.8,0.0) rectangle node{} (3,0.4);
\filldraw[thick, align=center, fill={rgb:white,5;pink,2}] (3,0.0) rectangle node{} (3.2,0.4);
\filldraw[thick, align=center, fill={rgb:white,5;pink,2}] (3.2,0.0) rectangle node{} (3.4,0.4);
\filldraw[thick, align=center, fill={rgb:orange,1;yellow,2;blue,3}] (3.4,0.0) rectangle node{} (3.6,0.4);
\filldraw[thick, align=center, fill={rgb:orange,1;yellow,2;blue,3}] (3.6,0.0) rectangle node{} (3.8,0.4);
\filldraw[thick, align=center, fill={rgb:orange,0;yellow,1;blue,2}] (3.8,0.0) rectangle node{} (4.0,0.4);
\draw [
    thick,
    decoration={
        brace,
        mirror
    },
    decorate
] (0,-0.1) -- (2.0,-0.1);
\node[] at (1.0,-0.45) {$D_1$};
\draw [
    thick,
    decoration={
        brace,
        mirror
    },
    decorate
] (2,-0.1) -- (4.0,-0.1);
\node[] at (3.0,-0.45) {$D_1$};
\node[] at (4.6,0.2) {vs.}; 
\filldraw[thick, align=center, fill={rgb:orange,1;yellow,2;pink,5}] (5.2,0) rectangle node{} (5.4,0.4);
\filldraw[thick, align=center, fill={rgb:orange,1;yellow,2;pink,5}] (5.4,0.0) rectangle node{} (5.6,0.4);
\filldraw[thick, align=center, fill={rgb:orange,1;yellow,2;pink,5}] (5.6,0.0) rectangle node{} (5.8,0.4);
\filldraw[thick, align=center, fill={rgb:orange,1;yellow,2;pink,5}] (5.8,0.0) rectangle node{} (6.0,0.4);
\filldraw[thick, align=center, fill={rgb:orange,1;yellow,2;pink,5}] (6.0,0) rectangle node{} (6.2,0.4);
\filldraw[thick, align=center, fill={rgb:orange,1;yellow,2;pink,5}] (6.2,0.0) rectangle node{} (6.4,0.4);
\filldraw[thick, align=center, fill={rgb:orange,1;yellow,2;pink,5}] (6.4,0.0) rectangle node{} (6.6,0.4);
\filldraw[thick, align=center, fill={rgb:orange,1;yellow,2;pink,5}] (6.6,0.0) rectangle node{} (6.8,0.4);
\filldraw[thick, align=center, fill={rgb:white,5;pink,2}] (6.8,0.0) rectangle node{} (7,0.4);
\filldraw[thick, align=center, fill={rgb:white,5;pink,2}] (7,0.0) rectangle node{} (7.2,0.4);
\filldraw[thick, align=center, fill={rgb:white,5;pink,2}] (7.2,0.0) rectangle node{} (7.4,0.4);
\filldraw[thick, align=center, fill={rgb:white,5;pink,2}] (7.4,0.0) rectangle node{} (7.6,0.4);
\filldraw[thick, align=center, fill={rgb:white,5;pink,2}] (7.6,0.0) rectangle node{} (7.8,0.4);
\filldraw[thick, align=center, fill={rgb:white,5;pink,2}] (7.8,0.0) rectangle node{} (8.0,0.4);
\filldraw[thick, align=center, fill={rgb:orange,1;yellow,2;blue,3}] (8.0,0.0) rectangle node{} (8.2,0.4);
\filldraw[thick, align=center, fill={rgb:orange,1;yellow,2;blue,3}] (8.2,0.0) rectangle node{} (8.4,0.4);
\filldraw[thick, align=center, fill={rgb:orange,1;yellow,2;blue,3}] (8.4,0.0) rectangle node{} (8.6,0.4);
\filldraw[thick, align=center, fill={rgb:orange,1;yellow,2;blue,3}] (8.6,0.0) rectangle node{} (8.8,0.4);
\filldraw[thick, align=center, fill={rgb:orange,0;yellow,1;blue,2}] (8.8,0.0) rectangle node{} (9.0,0.4);
\filldraw[thick, align=center, fill={rgb:orange,0;yellow,1;blue,2}] (9.0,0.0) rectangle node{} (9.2,0.4);
\draw [
    thick,
    decoration={
        brace,
        mirror
    },
    decorate
] (5.2,-0.1) -- (9.2,-0.1);
\node[] at (7.2,-0.45) {$D_2=2D_1$};

\end{tikzpicture}}
\caption{Toy example of CCDM amplitude sequences for 8PAM and probabilities $[0.4,0.3,0.2,0.1]$ for the four PAM amplitudes. Shaping block lengths $D_1=10$ (left) and $D_2=20$ (right). Colors identify amplitudes of PAM symbols.}
\label{f:CCDMexample}
\end{figure}
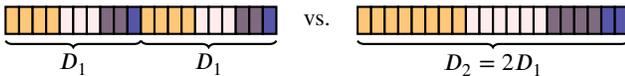
Attempts to shed more light on finite block-length effects for PAS with CCDM were made in \cite{fehenberger2019analysis}. First, it is noted that the mapping of amplitude-shaped PAM to QAM symbols has an effect on the moments of resulting QAM constellation for CCDM with finite block length. We will discuss this in more detail in Section~\ref{s:linearfiltermodel}. Second, and closer to the argument from \cite{dar2014shaping}, the limited concentration of identical symbols within a window of symbols is identified as an inherent property for short block-length CCDM. This is illustrated in Figure~\ref{f:CCDMexample}, which shows possible amplitude sequences for CCDM with $D_1=10$ and $D_2=20$ and four amplitudes with probabilities $[0.4,0.3,0.2,0.1]$. {Because CCDM with $D_1$ has shorter symbol run-lengths, we notice that the variation of energy across sliding windows of adjacent symbols is generally smaller than in CCDM with $D_2>D_1$. We make the notion of windowed energy sequences concrete in the following.}

The effects of moments and finite-length shaping can be captured to some extent by adjusting the EGN model. In addition to AMIN, the nonlinearity parameter $\eta$ also needs to incorporate the \textsl{channel memory}, which is a result of linear dispersion and walk-off in WDM transmission. Hence, akin to the finite-memory GN model in \cite{Agrell:2014}, a finite-memory EGN model is established. To this end, we consider windowed normalized energy 
sequences
\begin{equation}
\label{e:winseq}
e_k^w = \sum\limits_{i=k-\lfloor(w-1)/2\rfloor}^{i=k+\lfloor w/2\rfloor } \|x_i\|^2/E,
\end{equation}
where $E=\mathbb{E}(x_i^2)$ is the variance of the zero-mean data symbols and $w$ is the window length. In \cite{cho2022kurtosis}, energy sequences are used to compute {windowed} central moments
\begin{equation}
\label{e:wincent}
m_n^w = \frac{1}{w}\mathbb{E}\left[(e_k^w-w)^n \right],
\end{equation}
and the windowed standardized moments as 
\begin{equation}
\label{e:winkurt}
\begin{array}{rcl}
\mu_4^w &=& m_2^w + 1\\ 
\mu_6^w & =& m_3^w+3m_2^w+1
\end{array}\;.
\end{equation}
We note that the use of central moments in \eqref{e:wincent} is important as they capture the deviation around the average windowed energy. The NLIN from the average energy causes a constant phase offset and is thus compensated by a phase recovery at the receiver. In \cite{wu2021temporal}, $e_k^w$ is used to define the energy dispersion index (EDI) $\psi^w$, which is related to the second central moment by\footnote{If energy sequences are non-i.i.d.\ but wide-sense cyclostationary \cite{wu2021temporal}, then $m_n^w $ in \eqref{e:wincent} is the average over the cyclostationarity period.} 
\begin{equation}
\label{e:edi}
\psi^w = {E}\cdot m_2^w.
\end{equation}

\begin{figure}[t]
\includegraphics[width=0.45\textwidth]
{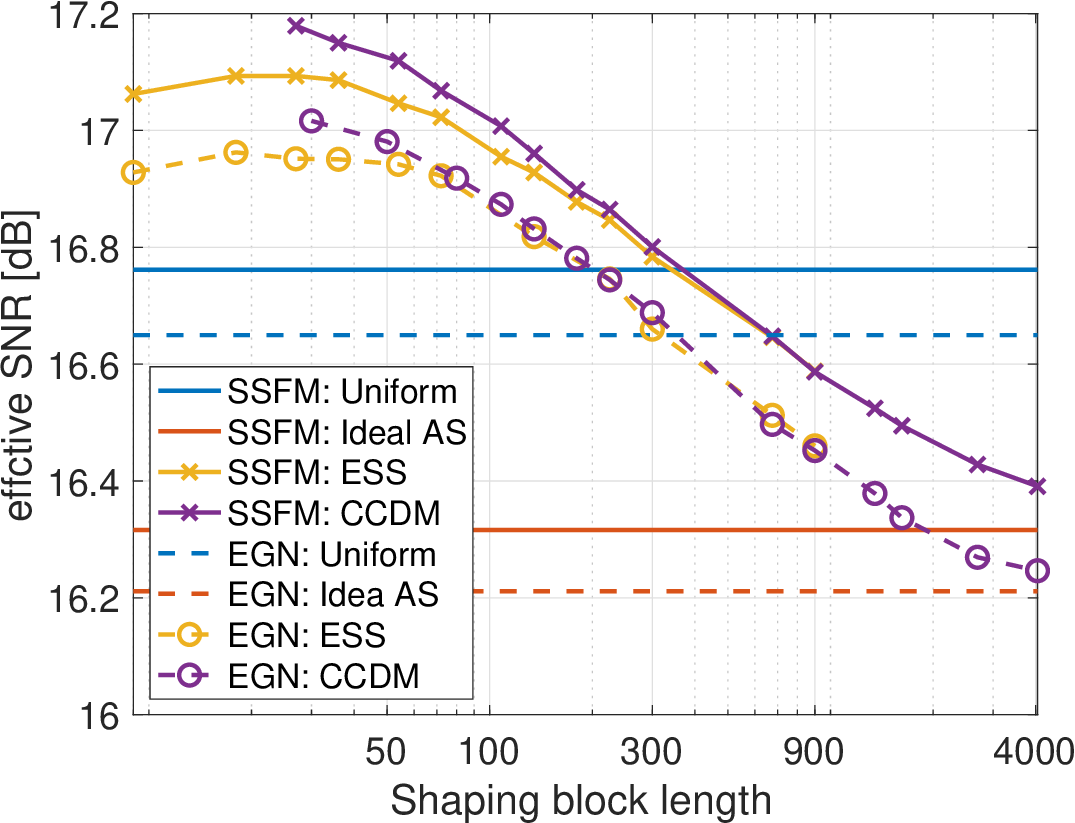}
\caption{Effective SNR versus shaping block length for ESS and CCDM. Simulation results using SSFM (``SSFM'')  as in Figure~\ref{f:SNReffective}. Analytical results using the implementation of the EGN model (``EGN'') from \cite[Appendix]{Dar:2014b} (assuming a pulse shape with zero roll-off) with the windowed versions of standardized moments in \eqref{e:winkurt}.}
\label{f:SNReffective2}
\end{figure}
The utility of the windowed moments can be seen in Figure~\ref{f:SNReffective2}, which shows the same simulation results as in Figure~\ref{f:SNReffective} (marked as ``SSFM'') together with analytical results using the implementation of the EGN model (marked as ``EGN'') from \cite[Appendix]{Dar:2014b}. We use the standardized moments $\mu_n$ for uniform transmission and shaping with infinite block length, and the windowed standardized moments $\mu_n^w$ for shaping with ESS and CCDM {with} finite block length. The window sizes are chosen as {\cite{cho2022kurtosis2,cho2022kurtosis}}
\begin{equation} 
{
\label{e:windowsizes}
\begin{array}{rcl}
\wspm&=&\lfloor 2\Rs \Bch |\beta_2|L \rceil\\
\wxpm&=&\lfloor 2\Rs \Bch \sqrt{\Nch \Bch /\Rs}|\beta_2|L\rceil
\end{array}}
\end{equation}
for self-phase modulation (SPM) and XPM induced NLIN, respectively. In \eqref{e:windowsizes},  $\Rs$, $\Bch$, $\Nch$, $\beta_2$, and $L$ are the baud rate, WDM channel spacing, number of WDM channels, group velocity dispersion, and link length, respectively. The expressions {are proportional to} the number of interfering symbols, i.e., the channel memory {\cite{Agrell:2014, wu2021temporal}}, due to dispersion within the channel of interest ($\wspm$) and walk-off between WDM channels ($\wxpm$), where one window size is used for the latter and the factor $ \sqrt{\Nch \Bch /\Rs}$ balances increased walk-off and reduced contributions from more {distant} channels.

We observe a fairly close match of analytical and simulated SNR results in Figure~\ref{f:SNReffective}. The slight offset in SNR between simulated and analytical curves can be attributed to the different pulse roll-offs. What we would like to highlight are the identical trends in all curves. This supports the inclusion of window functions to measure moments when transmitted symbols are non-i.i.d.\ as in the case of finite-length shaping. A similar observation has been made using the EDI in \cite{wu2021temporal}. In particular, for CCDM and $D\le w+1$, the EDI $\Psi^w$ from \eqref{e:edi} can be written as 
\begin{equation}
\label{e:edi2}
-\log(\psi^w) = -\log(D+1)+\log({3w}/[(\mu_4-1)E])
\end{equation}
 (see \cite[Eq.~(49)]{wu2021temporal}). This matches well the linear decrease of the effective SNR with increasing shaping block length in the log-log scale displayed for a large range of block-length values in Figure~\ref{f:SNReffective}. The mismatch for relatively small block lengths is an artifact of the simplicity of the EDI parameter. It has been mitigated through the introduction of the exponentially weighted EDI (EEDI) \cite{wu2021exponentially}, which can be interpreted as a special case of the filter model presented in the next section.

\subsection{Linear Filter Model}
\label{s:linearfiltermodel}

The use of energy sequences \eqref{e:winseq} in windowed moments and EDI can be refined by starting from the perturbative time-domain model \eqref{e:perturbation}. For this, we use two approximations. First, we use the pulse-matching condition $p=m+n$ \cite{Mecozzi:2000},\cite[Sec.~X]{Mecozzi:2012} to greatly reduce the number of perturbation coefficients. Defining $\j\gamma C_{m,n,s}\coloneqq\chi_{m,n,m+n,s}$, we have 
\begin{equation}
\label{e:perturbation2}
\Delta x_{k,0}\approx \j\gamma \sum\limits_{m,n,s} x_{m+k,0}x_{n+k,s}x^*_{m+n+k,s}C_{m,n,s}\;,
\end{equation}
which is typically used for perturbation-based nonlinearity compensation \cite{tao2011multiplier,gao2014reducing}. Secondly, we partition the distortion into two terms 
\begin{equation}
\label{e:perturbation3}
\begin{array}{rcl}
\Delta x_{k,0}&\approx& \j \gamma x_{k,0}\sum\limits_{n,s}|x_{n+k,s}|^2 C_{0,n,s}+\Delta' x_{k,0}\\
&=& \j \gamma E x_{k,0} \sum\limits_{n,s}e_{k-n,s} h_{n,s}+\Delta' x_{k,0}
\end{array}\;,
\end{equation}
where $\Delta' x_{k,0}$ includes all the terms from \eqref{e:perturbation2} with $m\neq 0$ and $(n\neq 0,s=0)$, 
$e_{k,s}= |x_{k,s}|^2/E$ is the normalized symbol-energy sequence, and $h_{n,s}= C_{0,-n,s}\in \mathbb{R}$. Since $h_{n,s}$ are real-valued and $\gamma$ is small in the regime of weak nonlinearity, the first term corresponds to NLIN-induced phase noise. Since its mean value is compensated at the receiver, we will consider the phase-noise NLIN term
\begin{equation}
\label{e:perturbation4}
d_{k,0}=\gamma E \sum\limits_{s}\left((e_s-1)\ast h_{s}\right)_k \;,
\end{equation}
where $(a\ast b)_k=\sum\limits_{n}a_{k-n}b_n$ denotes convolution. Noting that $C_{0,n,s}$ are the dominating perturbation coefficients, we interpret expression \eqref{e:perturbation4} as an approximation of NLIN components as the outputs of linear time-invariant (LTI) filters applied to symbol-energy sequences. This permits us to gain additional insight into the interaction between transmit signals from finite-length shaping and the optical fiber link properties. 

The filter point-of-view was adopted in \cite{peng2020transmission} to derive a 4-dimensional mapping from shaped PAM to dual-polarized QAM transmitted symbols and extended further in \cite{Taha:2023}. We first revisit some of the results from \cite{Taha:2023} on filter properties and then discuss nonlinearity tolerant PAS with CCDM and ESS. For brevity, we focus on SPM, i.e., filter $h_{0}$ in the following. Analogous findings apply for XPM and $h_{s\neq 0}$. 

\subsubsection{Filter properties}

\begin{figure}[t]
\includegraphics[width=0.45\textwidth]
{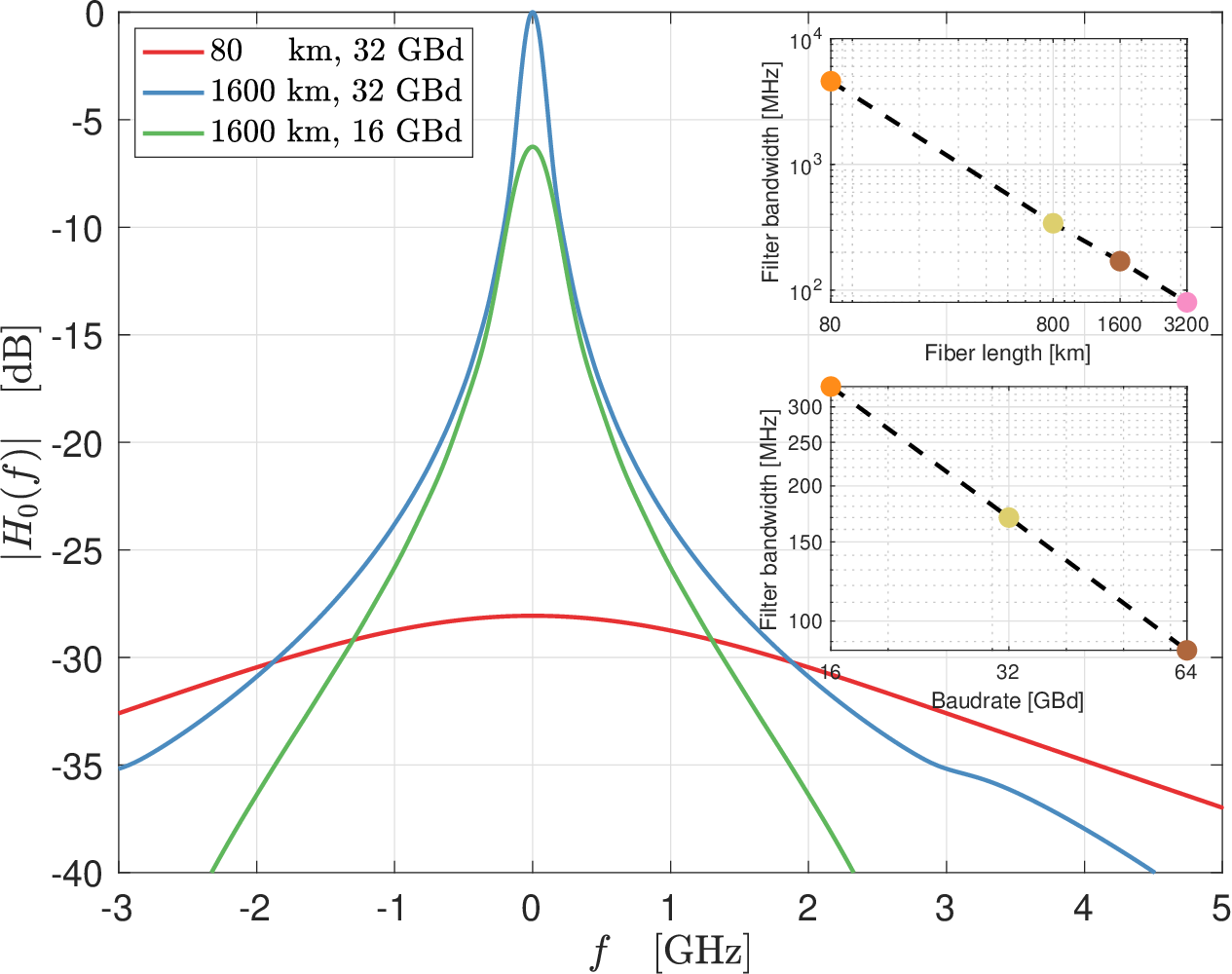}
\caption{Magnitude frequency response for SPM  LTI filter in \eqref{e:perturbation4},  for different link lengths and baud rates. Fiber parameters are as in Table~\ref{tab:setup1_sim_param}. Insets show 3~dB bandwidth of filters. }
\label{f:filterprop}
\end{figure}

As often done in LTI system analysis, we consider signals and systems in the frequency domain. Figure~\ref{f:filterprop} shows the magnitudes of frequency responses  $H_{0}$ of LTI filters for different link length and baud rate pairs. The fiber parameters are as in Table~\ref{tab:setup1_sim_param}. All results are normalized such that the $\max_f|H_{0}(f)|=1$ for the (1600~km, 32~GBd) case. We observe that the filters exhibit a lowpass characteristic. Comparing absolute values, the model confirms the absolute strength of NLIN increases with fiber length and baud rate. Similarly, the  lowpass properties become more pronounced with increasing link length and baud rate. This is highlighted in the two insets, which show the 3~dB filter bandwidth as a function of link length and baud rate, respectively. We observe a linear relationship in the log-log scale, which is consistent with the channel memory being proportional to $\Rs^2 L$, as shown in \eqref{e:windowsizes}.

\subsubsection{Shaping methods}

We now turn to the interaction of shaped signals and the NLIN filter. The strength of NLIN is represented by the power spectral density (PSD) of $d_{k,0}$ in \eqref{e:perturbation4}. Hence, we start by considering the autocorrelation $r_{ee}$ of the random process $(e_s-1)$. For i.i.d.\ signals, we have 
\begin{equation}
r_{ee,k}=(\mu_4-1) \delta_k,
\end{equation}
where $\delta_k$ is the Kronecker-delta function. For amplitude shaping with finite block length, the process $e_s$ is wide-sense cyclostationary. For CCDM, a closed-form expression for the average autocorrelation is given in \cite[Eq.~(34)]{wu2021temporal}. Accounting for the normalization of energy sequences with $E$ applied here, we obtain
\begin{equation}
r_{ee,k}=\left\{
\begin{array}{ll}
\mu_4-1,\;&k=0\\
(\mu_4-1)\left(\frac{|k|}{D}-1\right)\frac{1}{D-1},\;&1\le|k|<D\\
0,\;&|k|\ge D
\end{array}
\right.\;.
\end{equation}

\begin{figure}[t]
\centerline{%
\includegraphics[width=0.48\textwidth]
{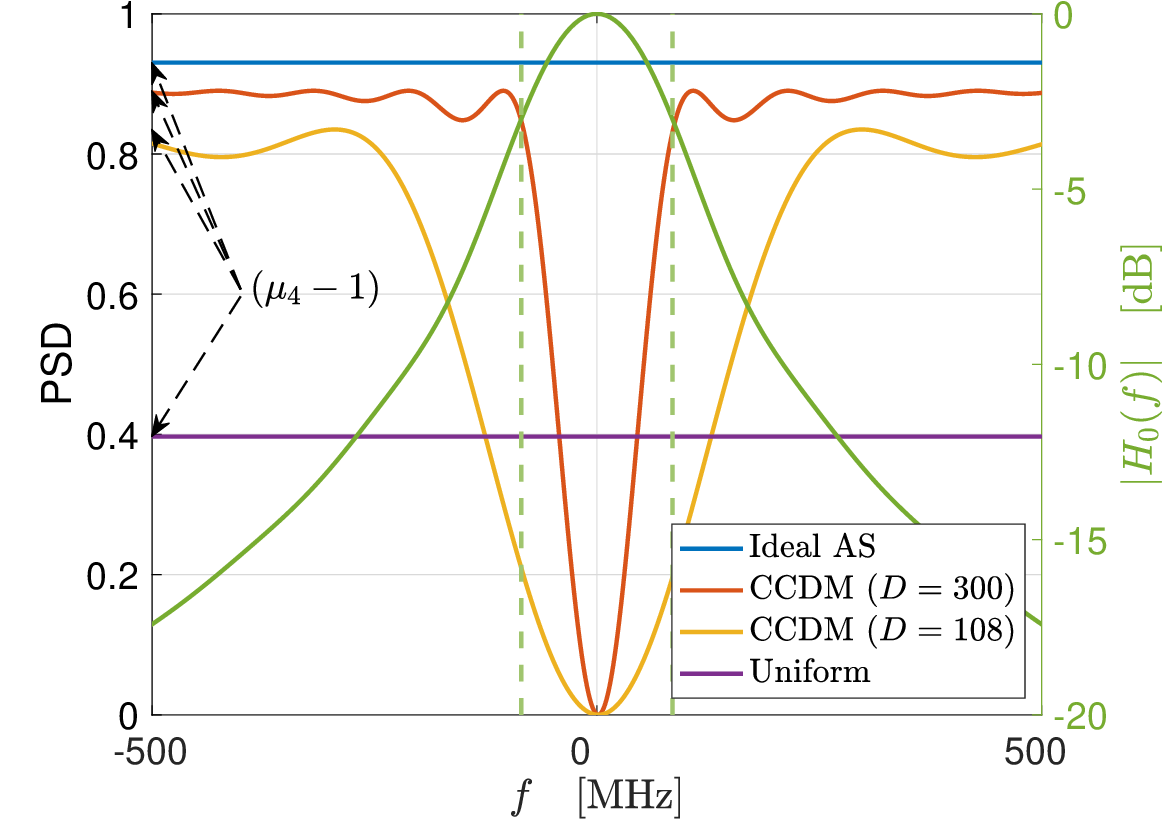}}
\caption{PSD (left axis) for energy sequences with different shaping methods. Overlaid (right axis) is the normalized magnitude frequency response for LTI filter $h_{0}$ in \eqref{e:perturbation4}, i.e., SPM. Dashed lines mark 3~dB bandwidth of filter. Parameters are as in Table~\ref{tab:setup1_sim_param}. }
\label{f:filterandESSCCDM}
\end{figure}

Figure~\ref{f:filterandESSCCDM} shows the PSDs of energy sequences for i.i.d.\ uniform and shaped transmission and for finite-length shaping with CCDM, with shaping parameters from Table~\ref{tab:setup1_sim_param}. The i.i.d.\ cases feature constant PSDs of value $\mu_4-1$, i.e., the effect of increased AMIN due to shaping is highlighted. For CCDM, we observe two effects that become more pronounced for shorter block lengths. First, the height of the PSD, which approaches $\mu_4-1$ for higher frequencies, reduces for shorter block length. Second, the PSDs exhibit a highpass characteristic. Referring to \eqref{e:perturbation4}, the area of the product of PSD and squared frequency response $|H_{0}|^2$ determines the strength of NLIN. Hence, considering the lowpass characteristic of $H_{0}$, as shown in Figure~\ref{f:filterprop} and overlaid  in Figure~\ref{f:filterandESSCCDM},  the improved nonlinearity tolerance of finite-length shaping can directly be inferred from the filter model. For example,  the lowpass property for optical fiber channels with significant memory together with the pronounced highpass property of energy sequences from CCDM with relatively short block length explain the occurrence of positive nonlinear shaping gains, as seen in terms of $\SNRnl$ in Figure~\ref{f:SNReffective2}.

\begin{figure}[t]
\centerline{%
\includegraphics[width=0.48\textwidth]
{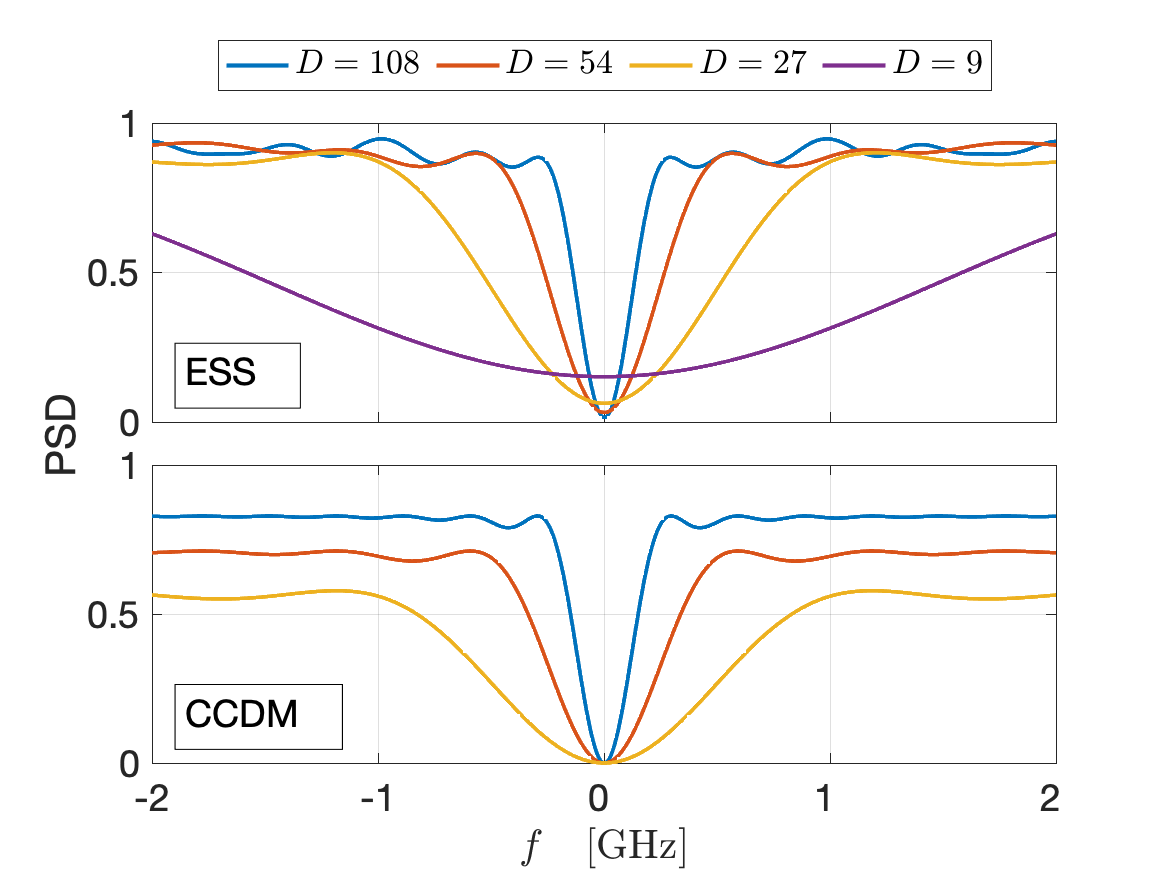}}
\caption{PSD for energy sequences with different shaping methods. Shaping parameters are as in Table~\ref{tab:setup1_sim_param}. }
\label{f:ESSCCDM}
\end{figure}

We can further compare the behaviour of finite-length CCDM and ESS considering the PSD plots in Figure~\ref{f:ESSCCDM}. 
For the case of ESS, we measured the empirical average autocorrelation to obtain the PSD. We note that the PSD for CCDM is exactly zero at frequency zero and that the notch widens with shorter block lengths. While the widening also happens for ESS, we observe that the PSD at $f=0$ increases as $D$ decreases. This is due to the non-constant energy of ESS-shaped sequences with larger variations for shorter block lengths. When considering this together with the lowpass characteristic of the fiber filter, we appreciate why $\SNRnl$ curves flatten and slightly decrease for ESS for very short block lengths in Figure~\ref{f:SNReffective2}. 

\subsubsection{Mapping}

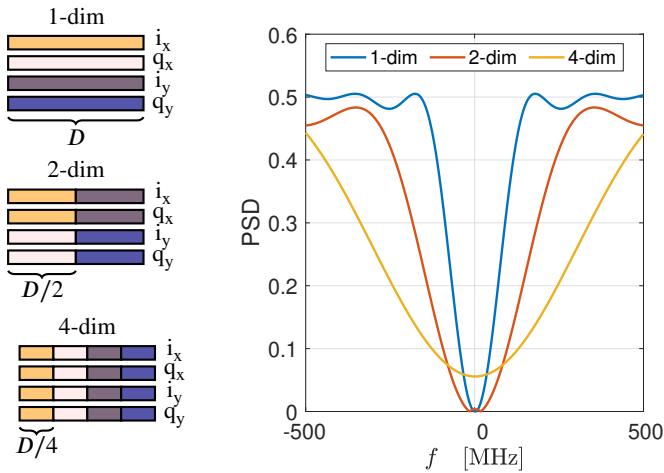
\begin{figure}[t]
\centerline{%
\parbox[][][t]{0.1\textwidth}{%
\begin{tikzpicture}[thick,scale=0.9, every node/.style={transform shape}]
\node[] at (4,1.4) {1-dim};
\filldraw[thick, align=center, fill={rgb:orange,1;yellow,2;pink,5}] (3,0.9) rectangle node{} (5,1.1);
\node[] at (5.3,1.05) {$\mathrm{i}_\mathrm{x}$};
\filldraw[thick, align=center, fill={rgb:white,5;pink,2}] (3,0.6) rectangle node{} (5.0,0.8);
\node[] at (5.3,0.725) {$\mathrm{q}_\mathrm{x}$};
\filldraw[thick, align=center, fill={rgb:orange,1;yellow,2;blue,3}] (3.0,0.3) rectangle node{} (5.0,0.5);
\node[] at (5.3,0.4) {$\mathrm{i}_\mathrm{y}$};
\filldraw[thick, align=center, fill={rgb:orange,0;yellow,1;blue,2}] (3,0) rectangle node{} (5.0,0.2);
\node[] at (5.3,0.05) {$\mathrm{q}_\mathrm{y}$};
\draw [
    thick,
    decoration={
        brace,
        mirror
    },
    decorate
] (3,-0.1) -- (5.0,-.1);
\node[] at (4.0,-0.4) {$D$};
\end{tikzpicture}

\begin{tikzpicture}[thick,scale=0.9, every node/.style={transform shape}]
\node[] at (4,1.4) {2-dim};
\filldraw[thick, align=center, fill={rgb:orange,1;yellow,2;pink,5}] (3,0.9) rectangle node{} (4,1.1);
\filldraw[thick, align=center, fill={rgb:orange,1;yellow,2;pink,5}] (3,0.8) rectangle node{} (4,0.6);
\node[] at (5.3,1.05) {$\mathrm{i}_\mathrm{x}$};
\filldraw[thick, align=center, fill={rgb:orange,1;yellow,2;blue,3}] (4,1.1) rectangle node{} (5.0,0.9);
\filldraw[thick, align=center, fill={rgb:orange,1;yellow,2;blue,3}] (4,0.8) rectangle node{} (5.0,0.6);
\node[] at (5.3,0.725) {$\mathrm{q}_\mathrm{x}$};
\filldraw[thick, align=center, fill={rgb:white,5;pink,2}] (3.0,0.5) rectangle node{} (4.0,0.3);
\filldraw[thick, align=center, fill={rgb:white,5;pink,2}] (3.0,0.2) rectangle node{} (4.0,0.0);
\node[] at (5.3,0.4) {$\mathrm{i}_\mathrm{y}$};
\filldraw[thick, align=center, fill={rgb:orange,0;yellow,1;blue,2}] (4,0.5) rectangle node{} (5.0,0.3);
\filldraw[thick, align=center, fill={rgb:orange,0;yellow,1;blue,2}] (4,0.2) rectangle node{} (5.0,0.0);
\node[] at (5.3,0.05) {$\mathrm{q}_\mathrm{y}$};
\draw [
    thick,
    decoration={
        brace,
        mirror
    },
    decorate
] (3,-0.1) -- (4.0,-.1);
\node[] at (3.5,-0.4) {$D/2$};
\end{tikzpicture}

\begin{tikzpicture}[thick,scale=0.9, every node/.style={transform shape}]
\node[] at (4,1.4) {4-dim};
\filldraw[thick, align=center, fill={rgb:orange,1;yellow,2;pink,5}] (3,0.9) rectangle node{} (3.5,1.1);
\filldraw[thick, align=center, fill={rgb:orange,1;yellow,2;pink,5}] (3,0.8) rectangle node{} (3.5,0.6);
\filldraw[thick, align=center, fill={rgb:orange,1;yellow,2;pink,5}] (3,0.5) rectangle node{} (3.5,0.3);
\filldraw[thick, align=center, fill={rgb:orange,1;yellow,2;pink,5}] (3,0.2) rectangle node{} (3.5,0);
\node[] at (5.3,1.05) {$\mathrm{i}_\mathrm{x}$};
\filldraw[thick, align=center, fill={rgb:white,5;pink,2}] (3.5,1.1) rectangle node{} (4.0,0.9);
\filldraw[thick, align=center, fill={rgb:white,5;pink,2}] (3.5,0.8) rectangle node{} (4.0,0.6);
\filldraw[thick, align=center, fill={rgb:white,5;pink,2}] (3.5,0.5) rectangle node{} (4.0,0.3);
\filldraw[thick, align=center, fill={rgb:white,5;pink,2}] (3.5,0.2) rectangle node{} (4.0,0);
\node[] at (5.3,0.725) {$\mathrm{q}_\mathrm{x}$};
\filldraw[thick, align=center, fill={rgb:orange,1;yellow,2;blue,3}] (4.0,1.1) rectangle node{} (4.5,0.9);
\filldraw[thick, align=center, fill={rgb:orange,1;yellow,2;blue,3}] (4.0,0.8) rectangle node{} (4.5,0.6);
\filldraw[thick, align=center, fill={rgb:orange,1;yellow,2;blue,3}] (4.0,0.5) rectangle node{} (4.5,0.3);
\filldraw[thick, align=center, fill={rgb:orange,1;yellow,2;blue,3}] (4.0,0.2) rectangle node{} (4.5,0.0);
\node[] at (5.3,0.4) {$\mathrm{i}_\mathrm{y}$};
\filldraw[thick, align=center, fill={rgb:orange,0;yellow,1;blue,2}] (4.5,1.1) rectangle node{} (5.0,0.9);
\filldraw[thick, align=center, fill={rgb:orange,0;yellow,1;blue,2}] (4.5,0.8) rectangle node{} (5.0,0.6);
\filldraw[thick, align=center, fill={rgb:orange,0;yellow,1;blue,2}] (4.5,0.5) rectangle node{} (5.0,0.3);
\filldraw[thick, align=center, fill={rgb:orange,0;yellow,1;blue,2}] (4.5,0.2) rectangle node{} (5.0,0.0);
\node[] at (5.3,0.05) {$\mathrm{q}_\mathrm{y}$};
\draw [
    thick,
    decoration={
        brace,
        mirror
    },
    decorate
] (3,-0.1) -- (3.5,-.1);
\node[] at (3.25,-0.4) {$D/4$};
\end{tikzpicture}
} \hfill
\parbox[][][t]{0.31\textwidth}{%
\includegraphics[width=0.31\textwidth]
{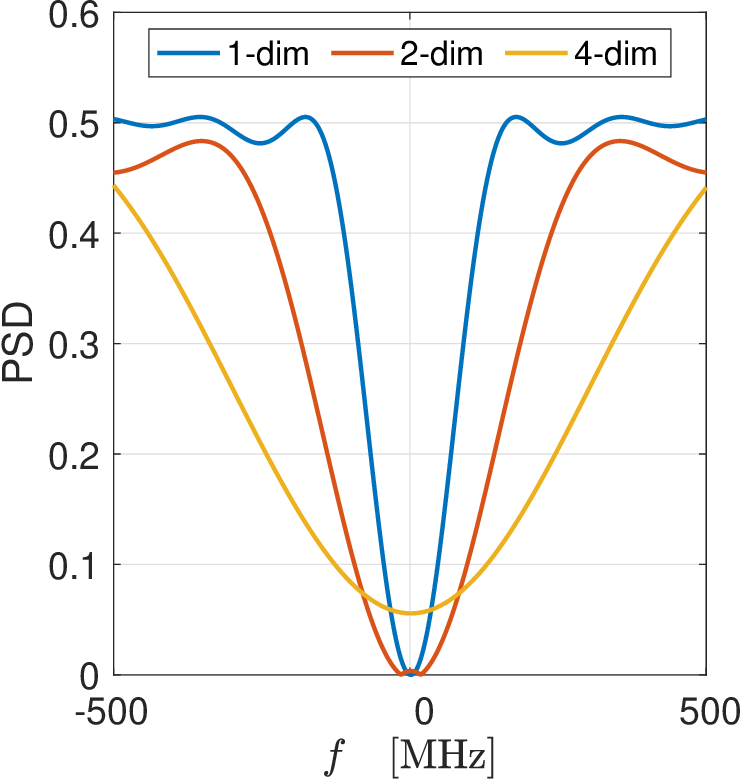}
}
}
\caption{Left: Illustration of 1-, 2-, and 4-dimensional mapping to in-phase (i) and quadrature (q) components of x- and y-polarization. Four shaping blocks of length $D$ are mapped to each $D$ QAM symbols in x- and y-polarization. Colors identify symbols from the same shaping block of length $D$. Right: PSD of energy sequences {$e_{\mathrm{agg},p}$} for CCDM with $D=180$ and the different mappings. Parameters are as in Table~\ref{tab:setup1_sim_param} but with dual polarization.}
\label{f:mapping}
\end{figure}

Different options for mapping from one-dimensional amplitude-shaped PAM symbol sequences to four-dimensional dual-polarized QAM symbols are illustrated in the left part of Figure~\ref{f:mapping}. The 1- and 2-dimensional mappings are also referred to as inter-DM and intra-DM pairing in \cite{fehenberger2019analysis}. {Extending the LTI filter analysis to a dual-polarized system involves examining the aggregated energy sequences \cite{Taha:2023}
\begin{equation}
\label{e:perturbationdual}
e_{\mathrm{agg},p} = 2e_p + e_{p' \neq p}
\end{equation}
where $e_p$ and $e_{p'}$  represent the energy of polarization $p$ and the orthogonal polarization $p' \!\! \neq \!\!p$, respectively.
The term $e_{\mathrm{agg},p}$ corresponds to the aggregated energy used to estimate the combined intra-polarization and inter-polarization NLI affecting the polarization 
$p$.}
The right-hand side of Figure~\ref{f:mapping} shows the PSDs for {$e_{\mathrm{agg},p}$} from CCDM with $D=180$ and different mappings. We note the lower height, indicative of moment $\mu_4$, and wider notch, indicative of finite-length effects, for 2-dim over 1-dim mapping. These characteristics suggest the use of 2-dimensional mapping for nonlinearity tolerant shaping and explain the corresponding observations  in  \cite{fehenberger2019analysis}.

The 4-dimensional mapping approach was chosen in \cite{peng2020transmission,Skvortcov:2020}
and demonstrated to further improve nonlinearity tolerance \cite{peng2020transmission,skvortcov2021huffman}. Considering the corresponding PSD shown in Figure~\ref{f:mapping}, we observe the wider notch but also an increased PSD around $f=0$. Accounting for the lowpass LTI filter as shown in Figure~\ref{f:filterprop}, we expect that 4-dimensional mapping has advantages for short fiber links with little lowpass characteristic, considered in \cite{peng2020transmission,skvortcov2021huffman}, but not for long fiber links. This has been verified through simulation results in \cite{Taha:2023}. 

\subsection{Carrier Phase Recovery}
\label{s:CPRmodel}
It is well known that part of the NLIN manifests as phase noise \cite{Mecozzi:2012,Secondini:2012, Dar:2013,Dar:2016}. 
This follows from \eqref{e:perturbation3} and noting that we operate in the weak nonlinearity regime. Accordingly, we approximate the signal plus interference as
\begin{equation}
\label{e:AMmodel}
\begin{array}{rcl}
x_{k,0}+\Delta x_{k,0}&\approx&x_{k,0}+\j x_{k,0}\gamma E  \sum\limits_{n,s}e_{k-n,s} h_{n,s}
+\Delta' x_{k,0}\\
&\approx&x_{k,0}\e^{\j \theta_k}+\Delta' x_{k,0}
\end{array},
\end{equation}
where $\theta_k=\gamma E  \sum_{n,s}e_{k-n,s} h_{n,s}$. From this expression, it is obvious that the carrier-phase recovery (CPR) included in coherent receivers may partially compensate NLIN \cite{Dar:2014}. 

We again employ the LTI filter model from the previous section to provide insights into the interplay of nonlinearity-tolerant shaping and CPR. For this, we adopt the CPR model from \cite{Civelli:2023}
\begin{equation}
\label{e:CPR}
\hat{\theta}_k = \sum\limits_{m=-\Ncpr}^{\Ncpr}c\cdot \theta_{m+k}+\xi_k,
\end{equation}
where $c=\frac{1}{2\Ncpr+1}$, and $\xi_k$ is the CPR noise.
The parameter $\Ncpr$ balances the agility of the CPR, which improves with decreasing $\Ncpr$, and the variance of the CPR noise, which decreases with increasing $\Ncpr$.
Applying \eqref{e:CPR} to the phase-noise term in \eqref{e:AMmodel}, we obtain the residual phase-noise NLIN
\begin{equation}
\label{e:cpr2}
\begin{array}{rcl}
\Delta d_{k,0}&=& \sum\limits_{n,s}\left(e_{k-n,s}  - \sum\limits_{m=-\Ncpr}^{\Ncpr} c\cdot e_{m+k-n,s}\right)h_{n,s}+\xi_k\\
&=&\sum\limits_s \left((e_s-1)\ast u\ast h_{s}\right)_k+\xi_k
\end{array}\;,
\end{equation}
where $u_m=\delta_m-c$ for $m=-\Ncpr,\ldots,\Ncpr$ and zero otherwise. We thus can model the CPR as another filter $u$ applied to the shaped energy sequence. {An expression that is equivalent to  \eqref{e:cpr2} is developed and used in the nonlinear phase noise metric in \cite{Civelli:2023}.}

\begin{figure}[t]
\centerline{%
\includegraphics[width=0.48\textwidth]
{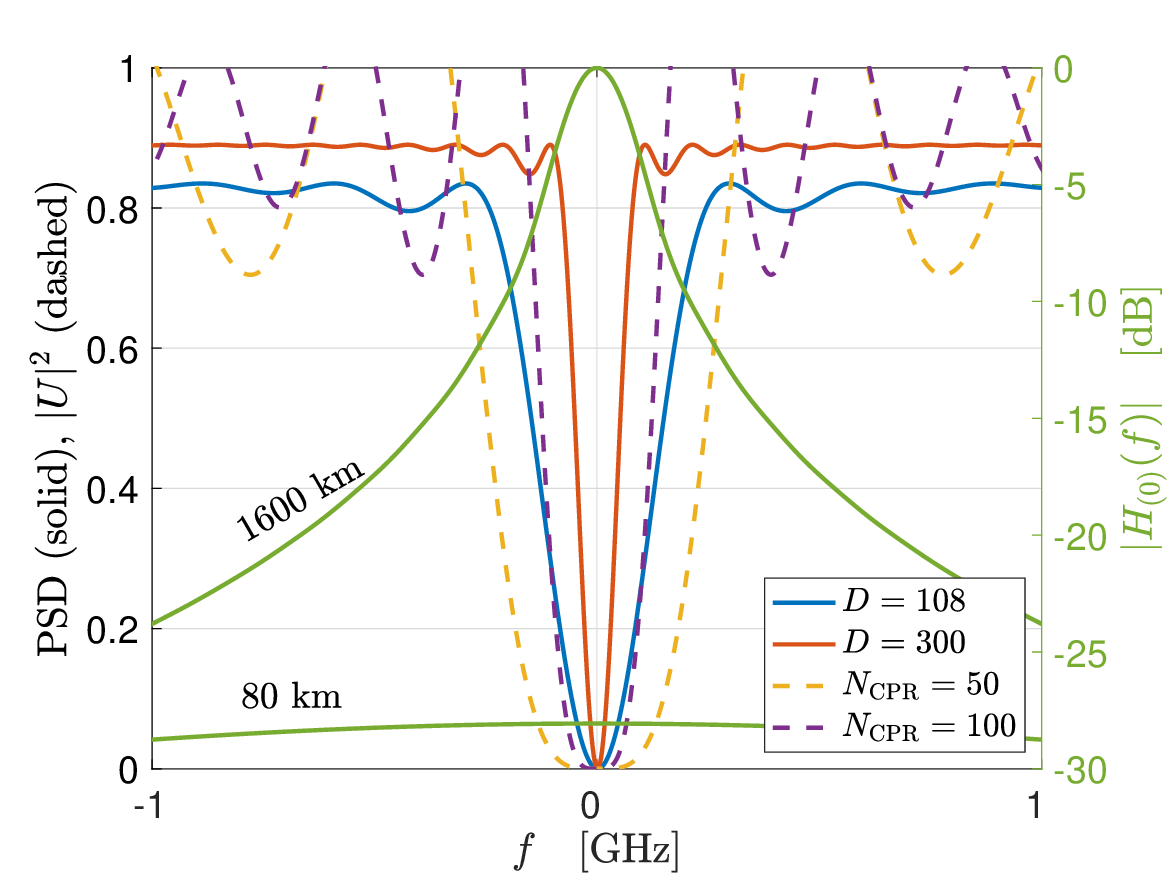}} 
\caption{
Left axis: (i) PSD for energy sequences shaped with CCDM and $D\in\{108,300\}$ and (ii) squared magnitude frequency response of CPR filter with $\Ncpr\in\{50,100\}$. Right axis: Normalized magnitude frequency response  for SPM  filter $h_{0}$ for link lengths $L\in\{80,1600\}$~km. 
Parameters are as in Table~\ref{tab:setup1_sim_param} and the additional case of a 1-span link. 
}
\label{f:filterandCPR}
\end{figure}

Figure~\ref{f:filterandCPR} illustrates the interaction of finite-length shaping, CPR, and the nonlinear fiber channel. We present the PSDs for shaped energy sequences using CCDM with $D=108$ and $D=300$ (as in Figure~\ref{f:filterandESSCCDM}), the magnitude frequency response of the CPR filter $u$ for $\Ncpr=50$ and $\Ncpr=100$, and the magnitude frequency response of the SPM filter $h_{0}$ with link lengths $L=1,600$~km and $L=80$~km. 
 We observe that the CPR suppresses low-frequency components and thus has a similar effect on NLIN as finite-length shaping. A smaller CPR window length corresponds to a wider notch and thus achieves better NLIN suppression. However, it also suffers from an increase in the CPR noise $\xi_k$, which cannot be seen in the figure.  Furthermore, the choice of $\Ncpr$ is dictated by the expected level of laser phase noise. As such, it is not a free parameter for the purpose of nonlinearity mitigation. These two aspects are decidedly different from nonlinearity tolerant shaping. The impact of the choices of $D$ for NLIN tolerance and $\Ncpr$ for mitigation are determined by the link setup as can  be seen from the overlaid frequency responses for the two link filters in Figure~\ref{f:filterandCPR}.

 We note that the filter point of view only considered part of the NLIN related to energy sequences, which can also be considered the multiplicative distortion  as per \eqref{e:AMmodel}. However, especially since CPR has the potential to mitigate the multiplicative distortion, the additive term also plays an important role in the context of nonlinearity tolerance through shaping. This will be elaborated on in the next section.

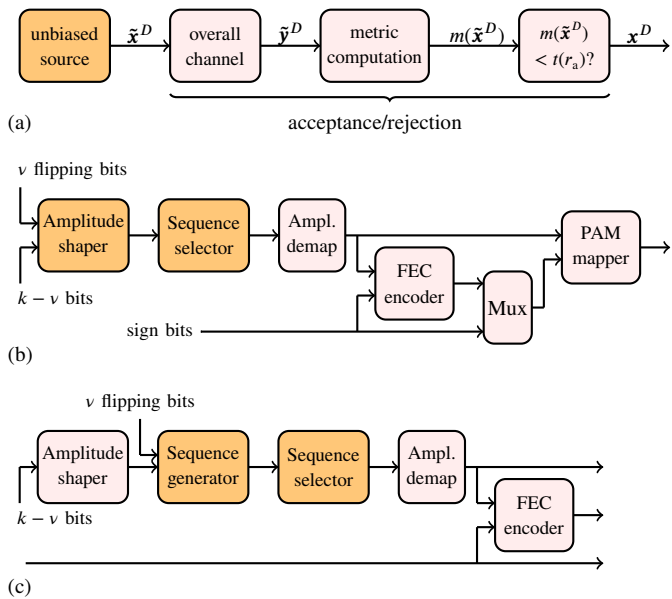
\begin{figure}[t]
\begin{tikzpicture}[thick,scale=0.8, every node/.style={transform shape}]
\filldraw[thick, align=center, fill={rgb:orange,1;yellow,2;pink,5}, rounded corners] (-3.8,0.2) rectangle node{\small unbiased\\[-.0em]\small source} (-2.3,1.4);
\node[] at (-3.8,-0.5) {(a)};
\draw[->] (-2.3,0.8) -- (-1.3,0.8);
\node[] at (-1.8,1) {$\ve{\tilde x}^D$};
\draw[thick, align=center, fill={rgb:white,5;pink,2}, rounded corners] (-1.3,0.2) rectangle node{\small overall \\[-.0em] \small channel} (0.2,1.4);
\draw[->] (0.2,0.8) -- (1.2,0.8) node[anchor=west] {};
\node[] at (0.7,1) {$\ve{\tilde y}^D$};
\draw [
    thick,
    decoration={
        brace,
        mirror
    },
    decorate
] (-1.3,0) -- (6,0);
\node[] at (2.1,-0.5) {acceptance/rejection};
\filldraw[thick, align=center, fill={rgb:white,5;pink,2}, rounded corners] (1.2,0.2) rectangle node{\small metric\\[-.0em]\small computation}(3,1.4);
\draw[->] (3,0.8) -- (4.5,0.8) node[anchor=west]{}; 
\node[] at (3.8,1) {$\m(\ve{\tilde{x}}^D)$};
\node[] at (3.6,0.5) {}; 
\filldraw[thick, align=center, fill={rgb:white,5;pink,2}, rounded corners] (4.5,0.2) rectangle node{\small $m(\ve{\tilde{x}}^D)$\\[-.0em]\small $<\t(\r)$?}(6,1.4);
\draw[->] (6,0.8) -- (7,0.8) node[anchor=west] {};
\node[] at (6.5,1) {$\ve{x}^D$};
\end{tikzpicture}

\vspace*{2mm}
\begin{tikzpicture}[thick,scale=0.8, every node/.style={transform shape}]
\node[] at (-4.8,-1.2) {(b)};
\draw[->] (-1.8,-0.8) -- (2.9,-0.8);
\draw[-] (-1.8,-0.8) -- (-1.8,-0.8) node[anchor=east] {\!\!\!\small sign bits};
\draw[->] (-4.8,0.6) node[anchor=east]{}-- (-4.5,0.6);
\draw[-] (-4.8,0.6) -- (-4.8,0) node[anchor=north  west] {\!\!\!\small $k-\nu$ bits};
\filldraw[thick, align=center, fill={rgb:orange,1;yellow,2;pink,5}, rounded corners] (-4.5,0.2) rectangle node{\small Amplitude\\[-.0em]\small shaper} (-3,1.4);
\draw[->] (-3,0.8) -- (-2.5,0.8);
\draw[->] (-4.8,1.0)  -- (-4.5,1.0);
\draw[-] (-4.8,1.0) -- (-4.8,1.6) node[anchor=south west] {\!\!\!\small $\nu$ flipping bits};
\filldraw[thick, align=center, fill=
{rgb:orange,1;yellow,2;pink,5}, rounded corners] (-2.5,0.2) rectangle node{\small Sequence\\[-.0em]\small selector} (-1,1.4);
\draw[->] (-1,0.8) node[anchor=east] {} -- (-.5,0.8);
\draw[thick, align=center, fill={rgb:white,5;pink,2}, rounded corners] (-.5,0.2) rectangle node{\small Ampl.\ \\[-.0em]\small demap} (0.6,1.4);
\draw[->] (0.6,0.8) -- (4.2,0.8) node[anchor=west] {};
\draw[-] (0.8,0.8) -- (0.8,0.2) node[anchor=west] {};
\filldraw[thick, align=center, fill={rgb:white,5;pink,2}, rounded corners] (1.1,-0.6) rectangle node{\small FEC\\[-.0em]\small encoder}(2.4,0.6);
\draw[->] (0.8,0.2) -- (1.1,0.2) node[anchor=west] {};
\draw[->] (0.8,-0.2) -- (1.1,-0.2);
\draw[-] (0.8,-0.8) -- (0.8,-0.2);
\draw[->] (2.4,0) -- (2.9,0);
 \draw[thick, fill={rgb:white,5;pink,2}, rounded corners] (2.9,-1.0) rectangle node{Mux} (3.7,0.2);
 \draw[-] (3.7,-0.4) node[anchor=east] {} -- (3.9,-0.4);
 \draw[-] (3.9,-0.4) node[anchor=east] {} -- (3.9,0.4);
 \draw[->] (3.9,0.4) node[anchor=east] {} -- (4.2,0.4);
   \draw[thick, align=center, fill={rgb:white,5;pink,2}, rounded corners] (4.2,0) rectangle node{\small PAM\\[-.0em]\small mapper} (5.5,1.2);
\draw[->] (5.5,0.6) -- (6,0.6) node[anchor=west]  {};
\end{tikzpicture}

\vspace*{2mm}
\begin{tikzpicture}[thick,scale=0.8, every node/.style={transform shape}]
\node[] at (-6.8,-1.2) {(c)};
\draw[->] (-6.7,-0.8) -- (2.9,-0.8);

\draw[->] (-6.8,0.8) node[anchor=east]{}-- (-6.5,0.8);
\draw[-] (-6.8,0.8) -- (-6.8,0.2) node[anchor=north  west] {\!\!\!\small $k-\nu$ bits};

\draw[->] (-4.8,1.0) -- (-4.5,1.0);
\draw[-] (-4.8,1.0) -- (-4.8,1.6) node[anchor=south] {\small $\nu$ flipping \small bits};
\filldraw[thick, align=center, fill={rgb:white,5;pink,2}, rounded corners] (-6.5,0.2) rectangle node{\small Amplitude \\[-.0em]\small shaper} (-5,1.4);
\draw[->] (-5,0.8) -- (-4.5,0.8);
\filldraw[thick, align=center, fill={rgb:orange,1;yellow,2;pink,5}, rounded corners] (-4.5,0.2) rectangle node{\small Sequence\\[-.0em]\small generator} (-3,1.4);
\draw[->] (-3,0.8) -- (-2.5,0.8);
\filldraw[thick, align=center, fill=
{rgb:orange,1;yellow,2;pink,5}, rounded corners] (-2.5,0.2) rectangle node{\small Sequence\\[-.0em]\small selector} (-1,1.4);
\draw[->] (-1,0.8) node[anchor=east] {} -- (-.5,0.8);
\draw[thick, align=center, fill={rgb:white,5;pink,2}, rounded corners] (-.5,0.2) rectangle node{\small Ampl.\ \\[-.0em]\small demap} (0.6,1.4);
\draw[->] (0.6,0.8) -- (2.9,0.8) node[anchor=west] {};
\draw[-] (0.8,0.8) -- (0.8,0.2) node[anchor=west] {};
\filldraw[thick, align=center, fill={rgb:white,5;pink,2}, rounded corners] (1.1,-0.6) rectangle node{\small FEC\\[-.0em]\small encoder}(2.4,0.6);
\draw[->] (0.8,0.2) -- (1.1,0.2) node[anchor=west] {};
\draw[->] (0.8,-0.2) -- (1.1,-0.2);
\draw[-] (0.8,-0.8) -- (0.8,-0.2);
\draw[->] (2.4,0) -- (2.9,0);
\end{tikzpicture}
\caption{(a) Illustration of the concept of sequence selection through generation from an unbiased source, i.e., not tuned to the shaping task, and rejection sampling. (b) Integration of sequence selection into PAS when retaining the amplitude shaper. $\nu$ redundant ``flipping'' bits are used to generate candidate sequences. (c) Decoupling of amplitude shaper and sequence generation and selection in PAS.}
\label{f:sequenceselection}
\end{figure}

\section{Sequence Selection}
\label{s:SS}
All shaping methods mentioned so far directly map data to shaped sequences. In this section, we deviate from this approach and permit the generation of a set of shaped sequences from which one sequence will be chosen for transmission, i.e., we consider sequence selection for shaping.

\subsection{Concept}
The concept of sequence selection was introduced as a means to estimate a capacity lower bound for optical fiber channels \cite{secondini2022new}. It has been implemented through rejection sampling as illustrated in Figure~\ref{f:sequenceselection}(a). For this, a source generates a sequence $\ve{\tilde x}^D$ of $D$ symbols drawn from some proposal distribution such as i.i.d.\ Gaussian \cite{secondini2022new}. To gauge its suitability for transmission, the fiber propagation and receiver processing such as dispersion compensation or single-channel backpropagation are emulated to obtain the corresponding sequence $\ve{\tilde{y}}^D$. Then, a metric $\m(\ve{\tilde x}^D)$ such as the $\ell^2$-norm between $\ve{\tilde{y}}^D$ and $\ve{\tilde{x}}^D$ is compared against a threshold $\t$, which is set according to an acceptance ratio $\r$. If the threshold test is passed, the sequence is accepted for transmission as $\ve{x}^D$.

\subsection{Integration into PAS}

\subsubsection{Direct approach and performance} Figure~\ref{f:sequenceselection}(b) shows the direct integration of sequence selection into PAS. A set of tentative sequences is generated by the amplitude shaper used in PAS. For example, CCDM has some natural redundancy if the shaping set of constant composition sequences is not a power of two. Additional redundant bits, so-called flipping bits \cite{wu2022list}, at the input of the amplitude shaper increase the flexibility to select from the shaping set. This is illustrated on the left-hand side of Figure~\ref{f:sequenceselection}(b). More redundant bits mean a larger set of candidate sequences but also a larger rate loss.  If we denote the number of input bits to the amplitude shaper to generate a sequence of $D$ amplitudes  by $k$, then the shaping rate is 
\begin{equation}
\label{e:shapingrate}
\Rshaper=\frac{k-\nu}{D},
\end{equation}
and thus the rate loss in \eqref{e:rloss} increases with $\nu$. Alternatively, we can interpret the role of $\nu$ as trading off linear and nonlinear shaping gains.

A second modification to PAS is the addition of the sequence selector in Figure~\ref{f:sequenceselection}(b). While the original sequence-selection framework  in \cite{secondini2022new} used the SSFM to mimic the optical fiber channel and $\ell^2$-norm between SSFM input and (processed) output to asses candidate sequences, the list-encoding CCDM (L-CCDM) proposed in  \cite{wu2022list} employs the 
EDI \eqref{e:edi} as a compact metric. Another metric that does not involve computationally complex SSFM is the lowpass-filtered symbol-amplitude sequence (LSAS) metric from \cite{Taha:2023}, which is based on the phase-noise NLIN term in \eqref{e:perturbation4}.

\begin{figure}[t]
\centerline{%
\includegraphics[width=0.45\textwidth]
{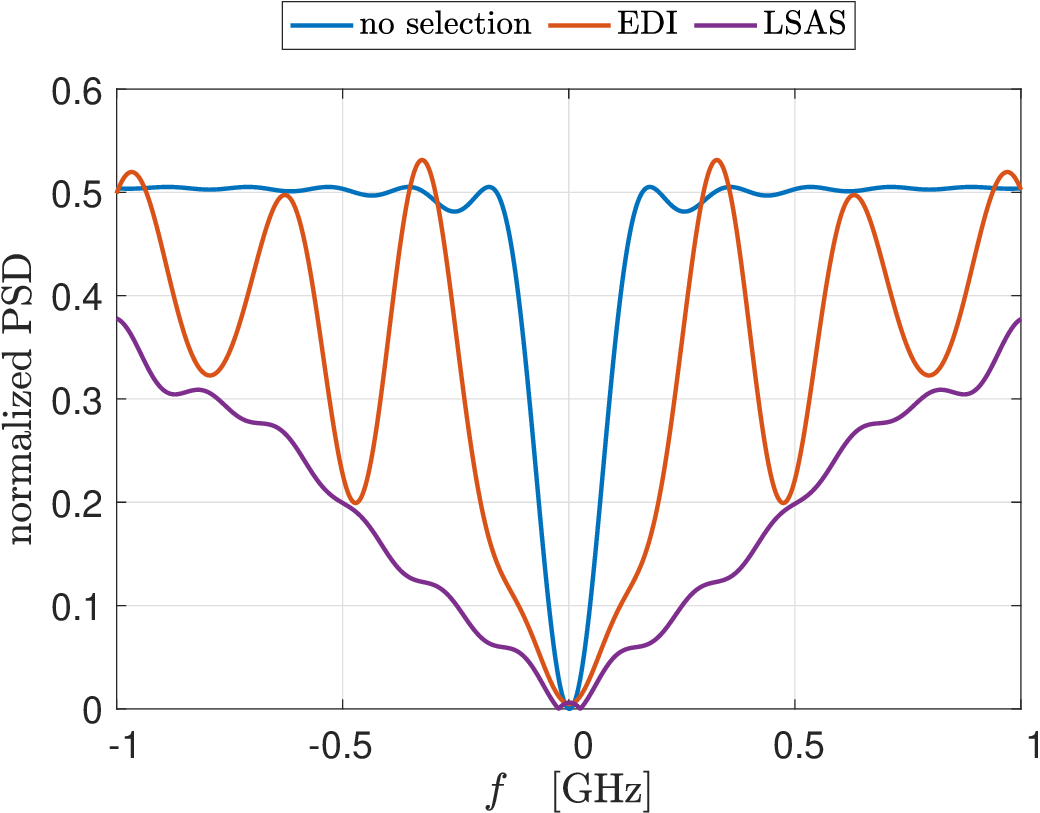}
}
\caption{PSD of energy sequences $\edp$ for CCDM with $D=180$: w/o sequence selection and with sequence selection using $\nu=2$ (16 candidate sequence) and EDI and LSAS metric. Parameters are as in Table~\ref{tab:setup1_sim_param} but with dual polarization.}
\label{f:selection}
\end{figure}
We can again consult the linear filter model from Section~\ref{s:linearfiltermodel} to gauge the effect of sequence selection on nonlinearity tolerance. Figure~\ref{f:selection} shows PSDs of shaped energy sequences using CCDM with $D=180$ and sequence generation with $\nu=2$, and selection with the EDI and LSAS metric, respectively. The example in the figure is for dual-polarized transmission with 1-dim mapping. This results in $4^{\nu=2}=16$ candidate sequences to select from.  We can observe how finite-length CCDM and sequence selection work together to shape sequences. The highpass characteristic that is associated with nonlinearity tolerance becomes more pronounced with sequence selection. We also see the benefit of the LSAS metric that is more directly associated with NLIN than the EDI metric. The latter leads to an oscillating PSD, which is an artifact of the windowing, i.e., rectangular filter, used in the EDI. 

\begin{figure}[t]
\center
\includegraphics[width=0.45\textwidth]
{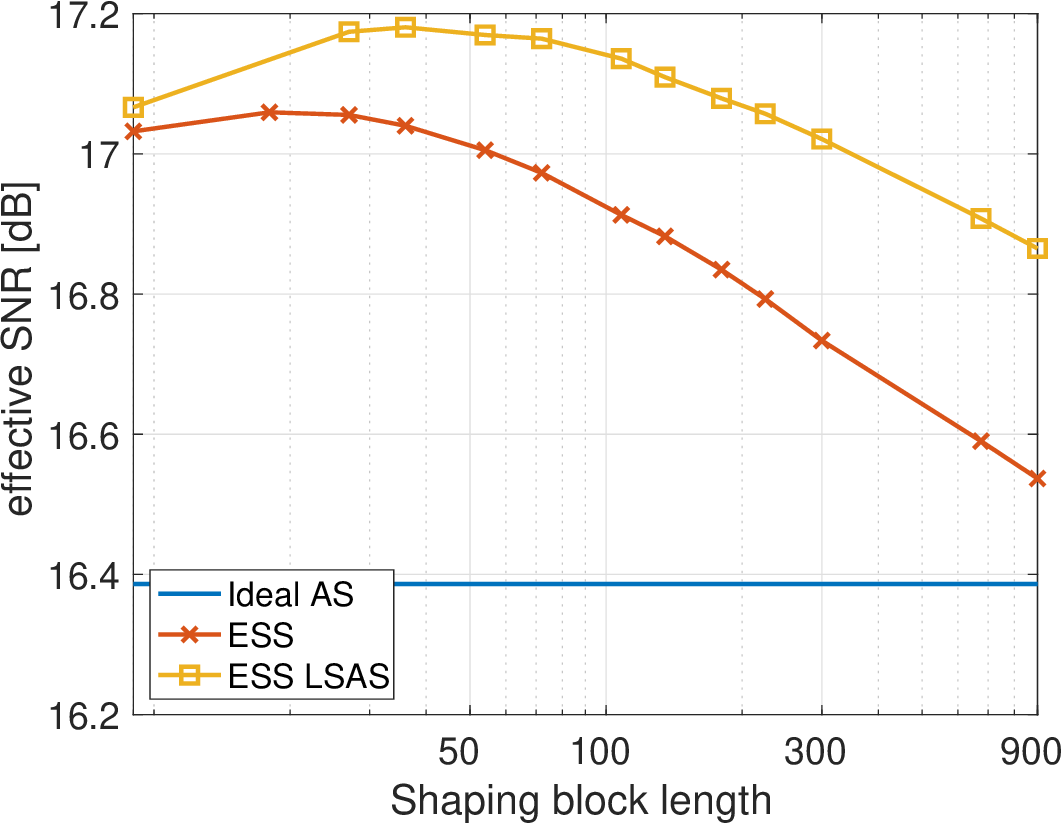}
\caption{Effective SNR versus shaping block length for ESS with and without sequence selection. Simulation results using SSFM for system parameters in Table~\ref{tab:setup1_sim_param} but with dual polarization.   $\nu=2$ and $4^\nu=16$ candidate sequences are available for sequence selection. LSAS metric is used for sequence selection. An LPA CPR with 1\% pilot rate is applied. Shaping with infinite block length
(“Ideal AS”) is included as a reference. }
\label{f:SNReffective3}
\end{figure}

Figure~\ref{f:SNReffective3} illustrates how sequence selection improves nonlinearity tolerance. We plot the effective SNR achieved for dual-polarized transmission and otherwise the parameters from Table~\ref{tab:setup1_sim_param} when using ESS with and without selection. Furthermore, we use a linear pilot-aided (LPA) CPR \cite{neshaastegaran2019log} with a 1\% pilot rate at the receiver, which mitigates NLIN as discussed in Section~\ref{s:CPRmodel}. {For simplicity, we chose the pilot symbols from the same shaped constellation as data symbols and adjusted the information rate to account for the pilots}. Consistent with the analysis in Figure~\ref{f:selection}, we observe a further increase of effective SNR when selection is added to finite-length shaping.




\subsubsection{Modifications}

The candidate generation using flipping bits and its integration into PAS as shown in Figure~\ref{f:sequenceselection}(b) has two disadvantages. First, the amplitude shaper is engaged for each candidate, which increases the computational complexity of the sequence selection scheme. Second, shaping and selection block lengths are necessarily identical. While both are influenced by the channel memory, other considerations such as rate loss and inter-block effects will determine their respective optimal values. Hence, they should be decoupled. 

The arrangement in Figure~\ref{f:sequenceselection}(c) overcomes those limitations. Consecutive outputs of the amplitude shaper can be concatenated to form one block for sequence selection. Then the flipping bits are applied to generate several candidate sequences. A simple but effective method is to perform symbol interleaving \cite{Civelli:2024}. The selected interleaver index, i.e., $\nu$ bits, need to be transmitted as side information. This can be accomplished using pilot symbols, as proposed in \cite{Civelli:2024}, or by FEC encoding the side information and using the resulting parity bits in place of sign bits, as suggested in \cite{taha2024perturbation}.

\subsection{Use of Sign Bits}

Practical sequence-selection metrics using EDI \cite{wu2022list} or LSAS \cite{Taha:2023} only account for the phase-noise NLIN term. As it can be seen in \eqref{e:AMmodel}, the multiplicative distortion term only depends on the amplitude of symbols, while the additive distortion term depends on both the energy and phase of symbols. Even though the multiplicative distortion is often the dominating component, this is not necessarily the case when a CPR is applied as discussed in Section~\ref{s:CPRmodel}. As pointed out in \cite{Civelli:2023,Civelli:2024}, performance gains with sequence selection may disappear when performing selection based on only symbol amplitudes. Therefore, also considering symbol phase, i.e., the sign bits produced by the FEC encoder in PAS, is desirable.


\subsubsection{Making sign bits available}

Using sign bits as part of shaping is a departure from the concept of PAS. In particular, it violates the concatenation of shaping and FEC coding shown in Figures~\ref{f:sequenceselection}(b) and~(c). Since sign bits are functions of the FEC codeword, selection of sign bits would need to be after the FEC, and the side information about the selection would not be protected by the FEC. However, it has been observed in  \cite{Civelli:2024} that knowing the sign bits for sequence selection is almost as effective as actually selecting sign bits. Hence, the structure in  Figure~\ref{f:sequenceselection}(c) could be retained if the sign bits that the FEC generates were available during amplitude-sequence selection.
To accomplish this, the bootstrap scheme from \cite{Boecherer2011} has been applied in \cite{Civelli:2024}. In this structure, the parity (i.e.\ sign) bits of each FEC code word are used as sign bits for the symbols associated with the next FEC code word. This is done for a number of code words. Following  \cite{Boecherer2011}, it is preferable to transmit the code words in reverse order so that they can be decoded at the receiver without delay. 

\subsubsection{Practical selection metrics}

A first practical sign-bit dependent sequence selection metric is the dispersion-aware EDI (D-EDI) proposed in  \cite{Liu:2024}. The D-EDI metric is the average of EDIs computed for the candidate sequence and several linearly dispersed versions of it corresponding to different propagation distances along the fiber. This partially accounts for the interaction of  phases and amplitudes of transmitted QAM symbols. The advantage of the D-EDI compared to using the SSFM is the significantly lower  computational complexity. {Another low-complexity approach to SSFM is the recently proposed coupled-band enhanced SSFM (CB-ESSFM), which employs subband processing to reduce the required number of SSFM steps, thereby significantly lowering computational complexity \cite{civelli2024cost}.}

An alternative metric that is more closely related to an analytical expression of NLIN as discussed in Section~\ref{s:linearfiltermodel} is a sign-bit dependent  extension of LSAS \cite{taha2024perturbation}. It is  based on the \textsl{additive-multiplicative (AM)} distortion model in \eqref{e:AMmodel} and reads 
\begin{equation}
\label{e:am_metric}
\metricam = \sum_{k} \bigg|\j \gamma E x_{k,0} \sum_{n,s}(e_{k-n,s}-1) h_{n,s} + \Delta' x_{k,0}\bigg|^2,
\end{equation}
where the outer summation is over all symbols within a candidate sequence. Extensions to dual-polarization and XPM distortions, relevant for digital sub-carrier multiplexing, are straightforward \cite{Taha:2023}. 

\subsubsection{Results}

We demonstrate the benefits of considering sign bits for sequence selection for the optical fiber communication system according to  Table~\ref{tab:setup1_sim_param} but with three WDM channels. We also include an LPA CPR with a pilot rate of 2.5\%  at the receiver. 

\begin{figure}[t]
\center
\includegraphics[width=0.45\textwidth]
{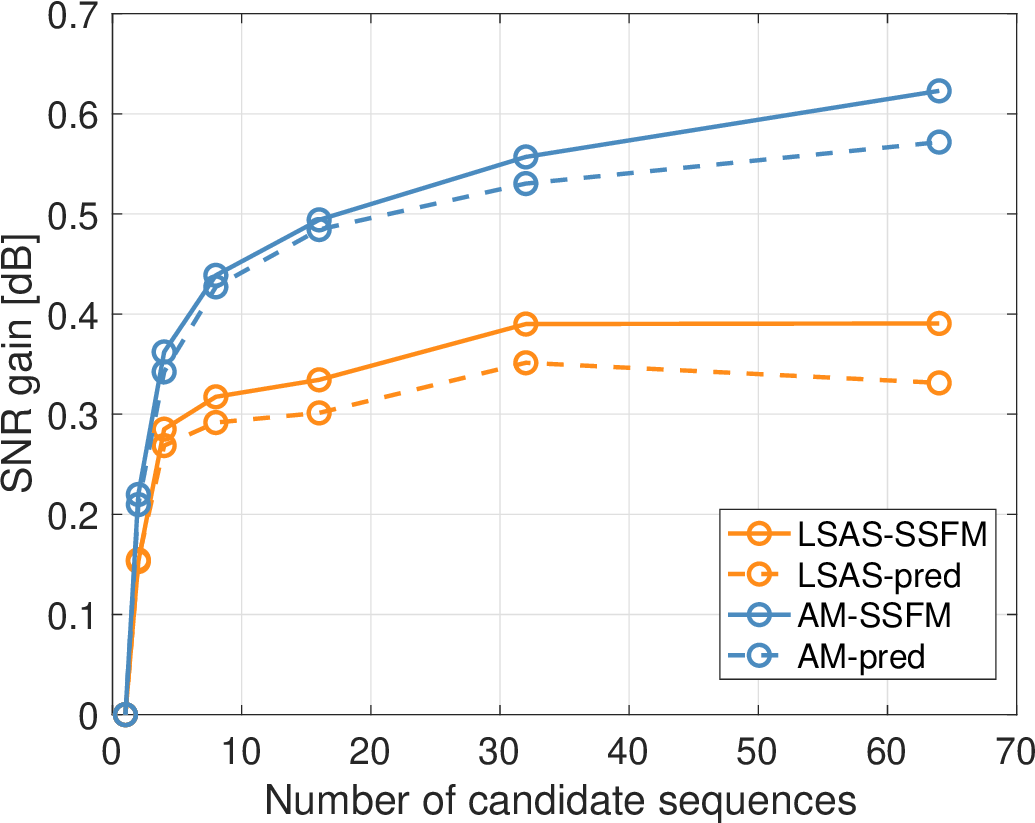}
\caption{Gain in effective SNR due to sequence selection versus number of candidate sequences. Simulation results using SSFM (solid lines) for system parameters in Table~\ref{tab:setup1_sim_param} but with three WDM channels. LSAS and AM metrics are used for sequence selection. An LPA CPR with a 2.5\% pilot rate is applied. Dashed lines are analytical results from \eqref{e:egngain}.}
\label{f:snr_pred}
\end{figure}

We first consider the asymptotic case of CCDM or ESS with a very long shaping block length and draw i.i.d.\ symbols from an MB distribution as input to sequence selection. As metrics we apply LSAS \cite{Taha:2023}, which only depends on symbol amplitudes, and the AM metric from \eqref{e:am_metric}. The sequences have a block length of 256. The SSFM simulation results for the gain in effective SNR as a function of the number of sequences  are shown as solid lines in Figure~\ref{f:snr_pred}. We can make two observations for this considered scenario. First, sequence selection based on only symbol amplitudes and thus phase-noise NLIN achieves an enhanced nonlinearity tolerance even when a CPR is applied. Second, accounting for sign-bits and thus the additive NLIN term provides another boost in tolerance, which for this particular test setup is an about 0.2~dB extra gain in effective SNR at $64$ candidates.

The AM model can also be used to predict the SNR gain provided by sequence selection.  Using \eqref{e:am_metric} for estimating $P_{\mathrm{NLIN}}$ and considering the effective SNR \eqref{e:SNReffopt} at optimum launch power in \eqref{e:PoptSNR}, we can express the SNR gain in dB as
\begin{equation}
 \label{e:egngain}\mathrm{SNR}^{\mathrm{opt}}_{\mathrm{sel}}- \mathrm{SNR}^{\mathrm{opt}}_{\mathrm{ref}} = \frac{1}{3}\left(P_{\mathrm{NLIN},\mathrm{ref}}-P_{\mathrm{NLIN},\mathrm{sel}}\right)\;,
\end{equation}
where $\mathrm{SNR}^{\mathrm{opt}}_{\mathrm{sel}/\mathrm{ref}}$ and $P_{\mathrm{NLIN},{\mathrm{sel}/\mathrm{ref}}}$ refer to the optimum effective SNR and the NLIN power corresponding to the systems with and without sequence selection, respectively. Figure~\ref{f:snr_pred} includes the predicted SNR gain as dashed lines.  The figure shows a fairly close match between the prediction using \eqref{e:egngain} and SSFM simulation results. This validates the use of the AM metric \eqref{e:am_metric} for sequence selection. It also suggests the use of \eqref{e:egngain} as a tool to guide the design of selection schemes, such as for determining the number of required candidates, without the need for SSFM simulations.  In this context, it is noted that the AM distortion model only assumes that a constant nonlinear phase-noise offset is compensated for via a CPR. It thus fails to fully account for the CPR effects. We speculate (i) that for this reason, the gain observed using SSFM simulation is slightly higher than the predicted gain in Figure~\ref{f:snr_pred}, and (ii) that sequence selection is only moderately affected by the presence of a CPR. 

\begin{figure}[t]
\center
\includegraphics[width=0.45\textwidth]
{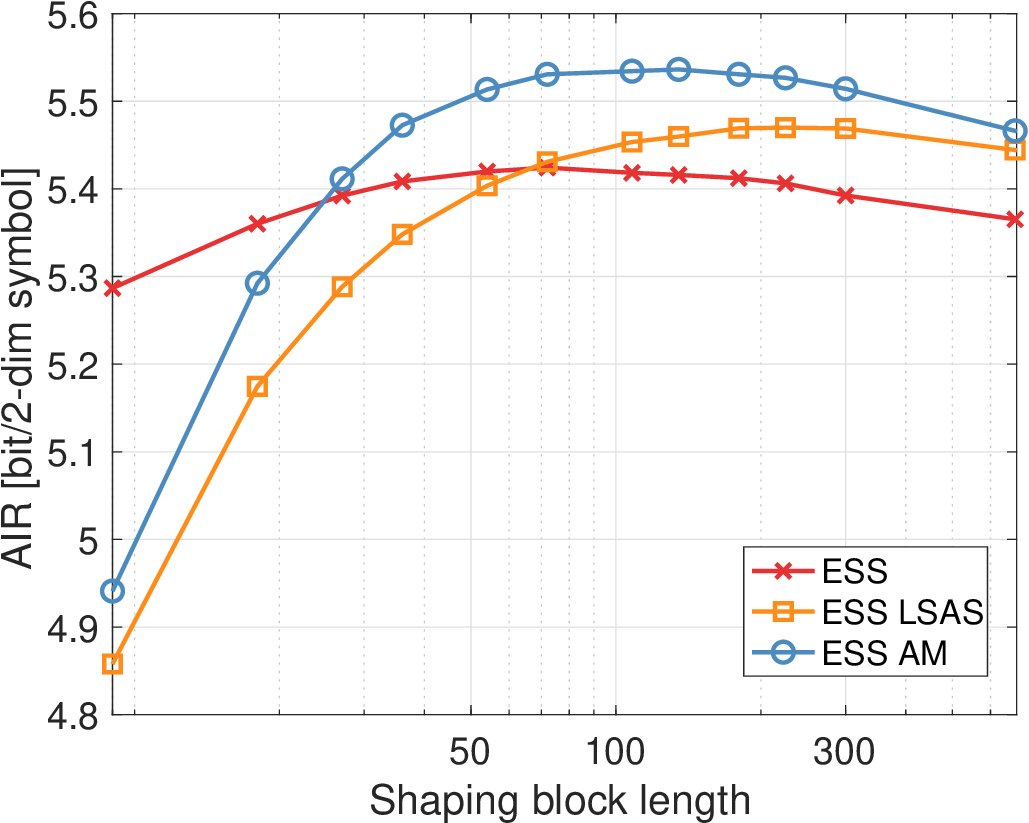}
\caption{AIR per two dimensions versus shaping block length. Simulation results using SSFM for system parameters in Table~\ref{tab:setup1_sim_param}.   $\nu=2$ and $4^\nu=16$ candidate sequences are available for sequence selection. LSAS and AM metric are used for sequence selection. An LPA CPR with 2.5\% pilot rate is applied.}
\label{f:air_am_lsas}
\end{figure}

Next, we perform sequence selection on each sequence generated by finite-length ESS, utilizing LSAS and AM metrics. Figure~\ref{f:air_am_lsas} illustrates the AIR as a function of shaping block length. Our observations are as follows. (i) For very short block lengths, the additional selection rate loss outweighs the nonlinear shaping gain, resulting in selection performance that is inferior to the baseline (i.e., ESS). This can be avoided by decoupling shaping and selection block lengths. (ii) Selection using the AM metric consistently outperforms LSAS, highlighting the importance of considering sign bits in practical finite length shaping scenarios. (iii) The AM metric yields an approximate gain of $0.08$~bit/2-dimenions in AIR (equivalent to a 0.4 dB gain in effective SNR) compared to the LSAS metric, with the optimal block length being significantly shorter than that of LSAS selection. This is advantageous for practical scenarios with low latency constraints.

{
\subsubsection{Computational Complexity}
Since candidate sequences can be efficiently generated using symbol interleaving, the primary computational burden in sequence selection arises from calculating metrics for each candidate sequence. We provide a brief analysis focusing on the number of terms used in the AM metric in \eqref{e:am_metric}. To accomplish this, we consider the expression of $\Delta x_{k,0}$ from \eqref{e:perturbation2}, noting that 
\begin{equation}
\label{e:AMmetricforComplexity}
\metricam=\sum_k\bigg|\Delta x_{k,0}-\j \gamma x_{k,0}E\sum_{n,s}h_{n,s}\bigg|^2,
\end{equation}
and we count the number of complex multiplications (CM) per QAM symbol required for metric evaluation. The computation of $\Delta x_{k,0}$ involves multiplying data-symbol triplets with perturbation coefficients $C_{m,n,s}$. Because the symbols $x_k$ are chosen from a square $M$-QAM constellation, the multiplications of the unsigned real and imaginary parts of the symbols can be precomputed and stored in a lookup table with $(\sqrt{M}/2)^3$ entries. As a result, the computational cost equals the number $N_{\mathrm{pb}}$ of perturbation coefficients. Additionally, we require two CMs to evaluate $\j \gamma x_{k,0}E\sum_{n,s}h_{n,s}$ and compute the squared magnitude. Therefore, the total computational cost of the AM metric is 
\begin{equation}
\CoAM = N_\mathrm{t}(2+\Npb),
\end{equation}
CMs per symbol, where $N_\mathrm{t}$ is the number of test candidates.

The number $N_{\mathrm{pb}}$ of perturbative coefficients $C_{m,n,s}$ used in \eqref{e:AMmetricforComplexity} depends on the channel memory. Considering the single-channel AM metric ($s=0$) for concreteness, and denoting the one-sided SPM memory measured in symbol intervals by $\wmem$, we observe that the sum in \eqref{e:perturbation2} is well approximated using only the terms with $|m| + |n| \le \wmem$. This constraint gives $\Npb = (\wmem+1)^2 + \wmem^2$ \cite{ARedyuk2020}, i.e., computational complexity grows quadratically with memory. However, several well-known complexity reduction techniques used for perturbation-based nonlinearity compensation can be applied, e.g., \cite{
Tao:2013,Li:2015,malekiha2016efficient,
ARedyuk2020}. First, a more economical coefficient selection approach employs the condition $|m\cdot n| < \wmem$. This leads to $\Npb = 4\sum_{k=1}^{\wmem} \lfloor (\wmem-1)/k \rfloor + 4\wmem + 1$ coefficients, which scales with $\wmem\log(\wmem)$ \cite{ARedyuk2020}. Additionally,  coefficient quantization has been applied, which groups coefficients with similar values into clusters and only uses cluster representatives in the AM metric. 
For instance, \cite{ARedyuk2020} applies K-means clustering to least-squares-adapted coefficents and found that $\Npb= 3 \lfloor (\wmem+14)/16\rfloor + 1$ is possible with only negligible performance loss. Given the high accuracy achieved with this heuristic for nonlinearity compensation, we will adopt it to quantify the number of distinct coefficients needed for AM metric evaluation with reduced complexity.
}
\begin{figure}[t]
\centerline{%
\includegraphics[width=0.45\textwidth]
{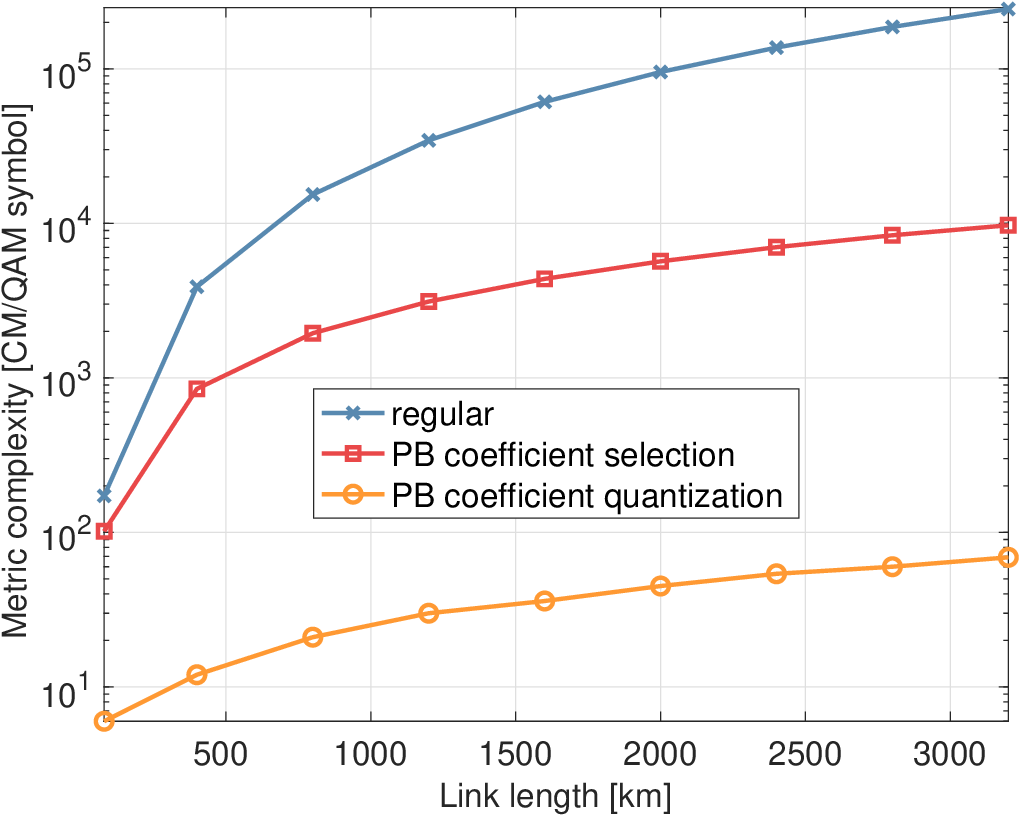}}
\caption{{Computational complexity for evaluating the AM metric versus link length. ``Regular'' uses all perturbation (PB) coefficients with $|m|+|n|\le \wmem$, ``PB coefficient selection'' applies $|m\cdot n|<\wmem$, and ``PB coefficient quantization'' uses the heuristic for K-means clustering from \cite{ARedyuk2020}.  System parameters from Table~\ref{tab:setup1_sim_param} with varying number of spans.}}
\label{f:complexity}
\end{figure}
{
Figure~\ref{f:complexity} illustrates the computational complexity for the various instances of $N_{\mathrm{pb}}$ as a function of fiber link length, based on the parameters in Table~\ref{tab:setup1_sim_param} and a varying number of spans. The one-sided SPM memory $\wmem$ is calculated using \cite[Eq.~(2)]{wu2021temporal}. 
We observe that complexity increases with the link length and corresponding channel memory. Calculating the AM metric without any complexity reduction method, referred to as ``regular'' in Figure~\ref{f:complexity}, results in prohibitively high computational complexity. However, using techniques to reduce the number of perturbation coefficients, in particular coefficient quantization, significantly lowers the computational burden by several orders of magnitude to a more acceptable level. 
}

\subsubsection{Extensions}
{
The mentioned previous works, aimed at reducing the complexity of nonlinearity compensation, have employed stochastic gradient descent \cite{malekiha2016efficient} and least-squares optimization \cite{ARedyuk2020} to learn perturbation coefficients. More recently, a similar learning approach has been applied specifically for accurate NLI modeling with fewer perturbation coefficients by employing normalized batch gradient descent \cite{barreiro2023data}. Interestingly, in addition to reducing computational complexity, learning the distortion model can potentially account for the effect of CPR.  
} This has been explored in \cite{barreiro2023phase} for a mean-phase rotation at the receiver. 

\begin{figure}[t]
\centerline{%
\includegraphics[width=0.45\textwidth]
{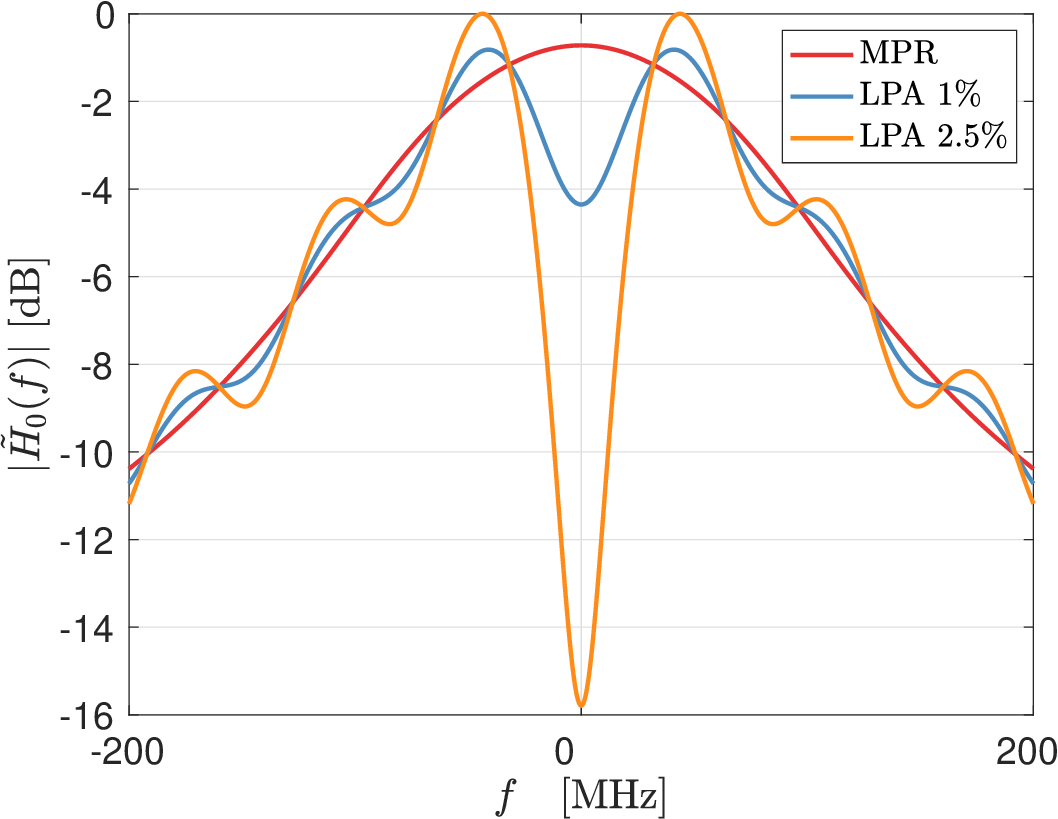}}
\caption{Normalized magnitude frequency response for learned overall LTI filter $\tilde h_{0}$, i.e., SPM, with three different CPR schemes. Parameters are as in Table~\ref{tab:setup1_sim_param}.}
\label{f:learned_filter}
\end{figure}

We consider this last point briefly further by returning to the filter model from Section~\ref{s:CPRmodel}. 
CPR-aware filter learning is equivalent to learning the overall filter 
\begin{equation}
{\tilde h}_{s} = u \ast h_{s}
\end{equation}
introduced in \eqref{e:cpr2}.  
 For this, we conduct an SSFM simulation of the system model from Table~\ref{tab:setup1_sim_param} without ASE noise, subtract the additive term $\Delta' x_{k,0}$ in \eqref{e:AMmodel} computed with numerical integration, and then learn the multiplicative coefficients in the filter ${\tilde h}_{s}$ using the recursive least-squares method.  Figure~\ref{f:learned_filter} shows the magnitude frequency response for the learned SPM filter $\tilde{h}_0$ for three different CPR methods: mean-phase rotation (MPR), LPA with 1\% pilots, and LPA with 2.5\% pilots. We observe that the result for MPR is consistent with the filter based on numerical integration and $u=\delta$ in \eqref{e:cpr2}. Different from this, the learned filters for the cases of LPA CPR experience a notch in the low-frequency region, which deepens as the pilot rate increases. This indicates the phase-noise tracking with increasing agility for higher pilot density, which also contributes to nonlinearity tolerance. 
 The results in Figure~\ref{f:learned_filter} also validate the conclusions drawn from the generic CPR model \eqref{e:CPR} used in Section~\ref{s:CPRmodel}.

\section{Conclusions and Remarks}
\label{s:conclusions}








In this paper, we provided a tutorial-style review of  the interplay between probabilistic shaping and fiber nonlinearity. We revisited the role of moments of the constellation that manifest in amplitude modulation induced noise and are well captured by the enhanced Gaussian noise model for the optical fiber channel. We explored the interaction of channel memory, fiber nonlinearity, and practical probabilistic amplitude shaping through time-domain perturbation analysis. 
The windowed-moment extension of the EGN and the linear filter model were identified as effective tools to express both AMIN and memory effects. Especially the latter supports an understanding of the nonlinearity tolerance of PAS with finite-length shaping methods and interactions with amplitude-to-symbol mappings, carrier-phase recovery, and sequence selection.    

An aspect of CPR that we did not cover but mention here in passing is that shaping can adversely affect decision-directed CPR schemes, such as blind phase search (BPS) \cite{mello2018interplay}. This is a result of transmitting low-energy symbols more frequently and needs to be addressed in the design of nonlinearity-tolerant schemes \cite{barbosa2020phase, askari2023bayesian, chimmalgi2023approximate}.
We have also not considered geometric constellation shaping, which moves from optimizing the distribution of constellation points to optimizing their location. This optimization is often considered in  the context of end-to-end learning \cite{Karanov:2018,Jovanovic:2023}. It can also be extended to joint optimization of geometric and probabilistic shaping  \cite{Aref:2022}, albeit without fitting into a concrete coded modulation structure such as PAS.

We consider sequence selection with easy-to-compute, sign-bit aware selection metrics as a favorable form of nonlinearity tolerant shaping. However, current solutions still suffer from a model and an algorithm deficit to take full advantage of nonlinearity tolerant transmission. Therefore, we believe that careful integration of learning-from-data concepts into a sequence-selection-PAS architecture is a meaningful path forward.

\bibliographystyle{IEEEtran}
\bibliography{references}

\end{document}